# Proximity coupling in superconductor–graphene heterostructures

Gil-Ho Lee[1,2] and Hu-Jong Lee[1]

[1]Department of Physics, Pohang University of Science and Technology, Pohang 790-784, Republic of Korea

[2]Department of Physics, Harvard University, Cambridge, Massachusetts 02138, USA

E-mail: lghman@postech.ac.kr (GHL) and hjlee@postech.ac.kr (HJL)

## Abstract

This review discusses the electronic properties and the prospective research directions of superconductor–graphene heterostructures. The basic electronic properties of graphene are introduced to highlight the unique possibility of combining two seemingly unrelated physics, superconductivity and relativity. We then focus on graphene-based Josephson junctions, one of the most versatile superconducting quantum devices. The various theoretical methods that have been developed to describe graphene Josephson junctions are examined, together with their advantages and limitations, followed by a discussion on the advances in device fabrication and the relevant length scales. The phase-sensitive properties and phase-particle dynamics of graphene Josephson junctions are examined to provide an understanding of the underlying mechanisms of Josephson coupling via graphene. Thereafter, microscopic transport of correlated quasiparticles produced by Andreev reflections at superconducting interfaces and their phase-coherent behaviors are discussed. Quantum phase transitions studied with graphene as an electrostatically tunable two-dimensional platform are reviewed. The interplay between proximity-induced superconductivity and the quantum-Hall phase is discussed as a possible route to study topological superconductivity and non-Abelian physics. Finally, a brief summary on the prospective future research directions is given.

# Contents

# 1. Introduction

Proximity coupling of a superconductor to graphene induces a unique coherent superconducting order in the graphene layer near the interface, which is often a result of the various special normal-state physical characteristics of graphene. This review will mainly focus on the experimental electrical transport studies of coherent proximity coupling in mesoscopic superconductor-graphene heterostructures, which have been extensively studied since the discovery of graphene. Some relevant theoretical studies are also reviewed. We begin with an introduction to the basic properties of graphene, with an emphasis on its physical interest and importance (Section 2). The graphene Josephson junction (GJJ) is discussed in

Section 3 among various other superconducting heterostructures. We then discuss the mesoscopic physics of different types of graphene–superconductor heterostructures, which again stems from the unique electronic properties of graphene (Section 4). Section 5 concludes this review and discusses some intriguing future research directions, including fundamental studies and application research. This review provides a summary of the development in graphene–superconductor heterostructures over the last decade and discusses exciting potential and the outlook of this field.

## 2. Basic properties of graphene

Graphene, a single atomic layer of carbon atoms in a hexagonal lattice structure, was the first true two-dimensional crystalline layer ever discovered. K. Novoselov and A. Geim were the first to isolate thin graphite and graphene out of bulk graphite [1]. Ever since its discovery, graphene has received enormous attention in the fundamental physics community and the industrial community, due to its high electrical conduction, high optical transparency, and extremely high mechanical strength.

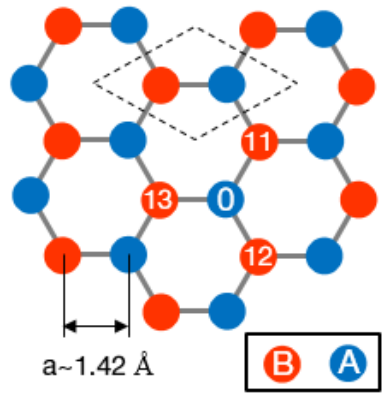

Figure 1. Hexagonal atomic structure of graphene. The dotted diamond represents the Bravais lattice consisting of A and B carbon atoms. The honeycomb lattice as a triangular Bravais lattice with a two-atom ('A' and 'B') basis can be seen.

Graphene has a Bravais lattice (dotted diamond) that consists of two identical carbon atoms, called the 'A' atom and the 'B' atom (Figure 1). The honeycomb lattice can be considered as two different triangular lattices, one with 'A' carbon atoms and the other with 'B' carbon atoms. The four valence electrons of each carbon atom (two from the 2s orbital and two from the 2p orbital) contribute to the bonding of the atomic structure. The other two electrons in the 1s orbital are well bounded and localized near the nuclei and hence, do not contribute to the low-energy transport properties of graphene. Three valence electrons in the sp$^2$-hybridized orbital contribute to three σ-bonding arms to link nearest-neighbor carbon atoms of a honeycomb lattice structure. The remaining electrons in the $p_z$ orbital globally interact (or conjugate) with each other to form π bonds. These π electrons are more loosely bound to the nuclei of carbon atoms in comparison to the other electrons and provide electronic transport in graphene.

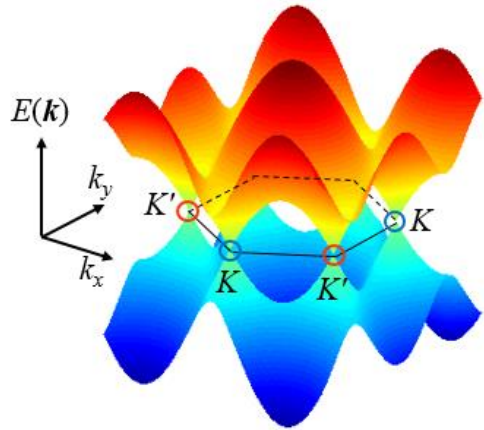

Figure 2. Electronic energy band structure calculated by the tight-binding model with the interactions from the nearest-neighbor atoms taken into account.

The electrons in the crystal experience a periodic potential produced by the arrangement of carbon atoms, the electronic wave functions of which can be described by the Bloch wave function. As shown in Figure 2, the tight-binding calculation up to first nearest-neighbor hopping gives a cone-like band structure, where the conduction band meets the valance band at a single point, named the Dirac point [2]. There are two distinctive Dirac points at two different valleys, K and K', and they correspond to the two equivalent carbon atoms 'A' and 'B' in a Bravais lattice. The linearity between energy and crystal momentum implies that the charge carriers behave as massless particles. In other words, the electronic structure of graphene is described by the relativistic Dirac equation even though all microscopic components describing graphene are nonrelativistic. While graphene itself is not intrinsically superconducting, it can inherit superconducting properties by being in proximity contact with a superconductor. Therefore, one has the unique opportunity to study the interplay of superconductivity and relativity in a real material. In addition, the electrostatic gate tunability of atomically thin graphene can provide fine control of the carrier density and the carrier types. Distinct from other semiconducting systems, the chemical inertness of graphene also provides a reliable and clean graphene/superconductor interface with a highly transparent superconducting contact. These extrinsic properties also make graphene an interesting platform to study proximity-induced superconductivity in heterostructures.

## 3. Graphene Josephson junction

Predicted by Josephson [3] in 1962 and experimentally confirmed a year later [4], the Josephson junction (JJ) was one of the earliest superconducting quantum devices that exploited the macroscopic quantum nature of superconductors. Early JJs were made from two superconducting metals separated by a thin insulating barrier where the Josephson coupling is mediated by Cooper pairs that tunnel through the barrier. Such tunneling JJs have been studied to investigate macroscopic quantum phenomena in condensed matter systems and, more recently, used for practical quantum devices such as superconducting quantum bits, the building block of quantum computers. Although the Josephson effect was originally studied in a tunneling JJ, any type of weak link between superconductors can exhibit dissipationless Josephson coupling through the proximity Josephson effect. Here, the insulating barrier is replaced by a non-superconducting spacer, such as a normal metal, a nano-constriction, or a semiconducting system such as an indium gallium arsenide (InGaAs) two-dimensional (2D) electron gas, carbon nanotubes, indium arsenide (InAs) nanowires, DNA, and so on. Recent developments in fabrication techniques have enabled a variety of different types of JJ to be realized. When a non-superconductor such as a normal metal (N) is in proximity contact with a superconductor, the superconducting correlation represented by the superconducting order parameter leaks into the N region with a characteristic superconducting coherence length $\xi_s$. This phenomenon is termed the superconducting proximity effect (Figure 3a). When two superconductors are closely placed such that the junction distance $L$ is comparable to or smaller than $\xi_s$ (Figure 3b), a proximity JJ is formed with dissipationless Josephson current flowing through the junction. In this superconductor–normal-metal–superconductor (SNS) proximity junction, conduction electrons mediate the pair-current transport from the left superconductor ($S_{left}$) to the right one ($S_{right}$) either by ballistic or diffusive transport through the N region.

In this section, we review seminal studies on the superconductor–graphene hybrid structure and superconductor–graphene–superconductor Josephson junctions (GJJs). When a mesoscopic conductor mediates the Josephson coupling, it is important to understand the relevant length scales for the proximity JJs. Various theoretical approaches for describing GJJs in relation to the length scales are examined in Section 3.1. The experimental aspects of realizing GJJs and the advancement of nano-fabrication techniques are reviewed in Section 3.2. The Josephson current $I_J(\phi)$ varies with the macroscopic phase difference, $\phi$, between the two superconductors. This current-phase relationship contains valuable information on the nature of the proximity JJ, such as the topological properties of the superconductivity or length scales of the coupling mechanism. The current-phase relationship of GJJs is discussed in Section 3.3. Lastly, we focus on the escaping dynamics of a phase particle studied in GJJs in Section 3.4.

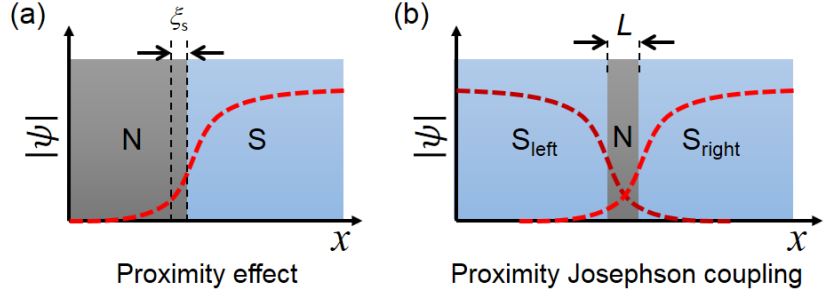

Figure 3. (a) The superconducting order parameter $|\Psi|$ of a superconductor (S) penetrating into the normal metal (N) with a length scale of the superconducting coherence length, $\xi_s$. (b) Order parameters from two sides have an overlap in N, producing proximity Josephson coupling.

1. Theoretical development

The mesoscopic GJJs are classified by comparing their junction length, $L$, with relevant length scales. The mean free path ($l_{mfp}$) of the graphene is the criterion for defining the ballistic and diffusive junction regimes, while the superconducting coherence length ($\xi_s$) separates the short and long junction regimes. Therefore, there can be at least four different junction regimes; short-ballistic, short-diffusive, long-ballistic, and long-diffusive. Since many experimental and theoretical studies on GJJs consider different junction regimes, we first summarize the relevant length scales to avoid confusion. For ballistic graphene, $\xi_s$ is given in terms of the reduced Planck constant, $\hbar$, the Fermi velocity in the graphene, $v_F$, and the superconducting gap of the electrode, $\Delta$, as $\xi_s = \hbar v_F/\Delta$. This describes the length scale of the superconducting correlation propagating from the superconductor into graphene. For diffusive graphene, this is replaced by by $\xi_s = \sqrt{\hbar D/\Delta}$ with a diffusion constant $D = v_F l_{mfp}/2$.

It is also possible to compare the energy scales of the system instead of considering the length scales. The relevant energy scale is the Thouless energy, $E_{Th}$, which is related to the dwell time, $\tau_d$, of quasi-particles in the junction, *i.e.* $E_{Th} = \hbar/\tau_d$. Naturally, it can then be given that $E_{Th} = \hbar v_F/L$ for ballistic systems and $E_{Th} = \hbar D/L^2$ for diffusive systems. As the smallest energy scale determines the behavior of the system, the properties of the system in the short junction regime, where $\Delta < E_{Th}$ (or equivalently, $L < \xi_s$), are determined by the superconducting gap. For the long junction regime of $\Delta > E_{Th}$ (or equivalently, $L > \xi_s$), however, the Thouless energy mainly governs the junction behavior.

Before any experimental investigations on GJJs and graphene–superconductor heterostructures, Titov and Beenakker theoretically studied Josephson coupling via graphene by solving the Dirac-Bogoliubov-de Gennes (DBdG) equations [5]. They focused on a wide ($W \gg L$) short-ballistic GJJ with "ideal" graphene–superconductor interfaces for conventional s-wave superconductivity. Although graphene by itself is not superconducting, the region in contact with a superconducting electrode is considered to acquire a non-vanishing superconducting order. The role of an s-wave superconductor is to couple electron-like and hole-like excitations with opposite spin and momentum, the latter of which means the pair carriers are in opposite valleys of the graphene. Titov and Beenakker calculated the Andreev bound states between two superconducting contacts and obtained the Josephson current. They predicted a finite Josephson current, even at the Dirac point where the nominal density of states in graphene vanishes. Surprisingly, $I_c$ and $R_N$ scale with $1/L$ and $L$, respectively, instead of being independent of $L$, as expected for short-ballistic junctions. This unusual "pseudo-diffusive" (or quasi-diffusive) behavior is closely related to the finite conductivity of graphene ($\sim 4h/\pi e^2$) appearing at the Dirac point, even without any impurities or scattering [6-8].

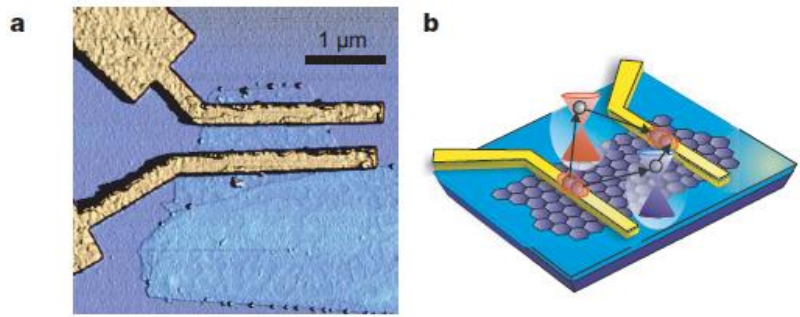

Figure 4. (a) Atomic-force microscope image of a monolayer-graphene Josephson junction (JJ). (b) The two time-reversed electrons which compose a Cooper pair enter into different valleys, represented by the red and blue cones [9].

The first experimental study on a GJJ showed strong gated modulation of the Josephson current and a bipolar nature by tuning the Fermi energy across the Dirac point (Figure 4) [9], as theoretically predicted in Ref. [5]. This GJJ was a long-diffusive junction ($l_{mfp}$, $\xi_s < L$) according to the estimation from the conductance given by the authors ($\mu$ = 1400 cm$^2$/Vs, $l_{mfp}$ = 20 nm, and $\xi_s$ = 250 nm). Nonetheless, it showed $I_cR_N$ to be a product of ~ $\Delta/e$, the size of which was comparable to the theoretically predicted value of ~ 2.5$\Delta/e$ for short-ballistic GJJs. However, it was not made clear if one can directly compare the magnitude of the measured Josephson current with theoretical prediction since the thermal fluctuations, quantum fluctuations, and the noise from the electromagnetic environment can reduce the measured Josephson current (switching current). This is further discussed in Section 3.4 as a separate topic. Many experiments, including this work, routinely display a non-vanishing Josephson current at the Dirac point, which may be a consequence of the characteristic properties of Dirac chiral fermions in two dimensions. However, impurity-induced residual carriers remaining at the Dirac point can also give similar results. Although recent experimental work with suspended GJJ [10] may have shown qualitative signatures of pseudo-diffusive behavior, a definitive experimental demonstration is still out of reach, mostly due to impurities in graphene and non-ideal superconducting contacts.

At a similar time, Moghaddam and Zareyan [11,12] solved DBdG equations for a short-ballistic GJJ with arbitrary $W$, and showed that the $I_c$ dependence on $W$ is very different for different edge configurations. For smooth and armchair edges, $I_c$ linearly increases with $W/L$ for large $W/L$ near the Dirac point, which reproduces the pseudo-diffusive behavior predicted by Titov and Beenakker. The absence of $I_c$ quantization is attributed to the fact that the evanescence modes dominate over the few quantized propagating modes near the Dirac point where the Fermi wavelength diverges. For zigzag edges, $I_c$ is quantized to $(n + 1/2) \times 4e\Delta/h$ with integer $n$ and resembles the conductance quantization of zigzag graphene nanoribbons and the half-integer quantum-Hall effect. It should be noted that experimentally it is very challenging to control the edges of graphene.

Besides the Josephson current, the subharmonic gap structure at a finite voltage has been calculated for short-ballistic GJJs by Cuevas and Yeyati [13]. This structure consists of a series of differential conductance maxima at voltages of $2\Delta/ne$ with integer $n$, which is the manifestation of multiple Andreev reflections (MARs) [14]. The subharmonic gap structure was suggested as a better comparison between experiment and theory as it is more robust against intrinsic and extrinsic fluctuations in comparison to the Josephson current. Although the MAR features are predicted to have a strong gate dependence for short-ballistic GJJs, this prediction remains to be verified experimentally. Most of the MAR features have been studied in the diffusive regime [9,15-17]. Du, Skachko, and Andrei [15] showed that the gate dependence of MAR follows a theoretical prediction for diffusive junctions. These initial theoretical and experimental works sparked significant interest in GJJ heterostructures and induced various theoretical approaches. In the aforementioned theoretical studies, fixed boundary conditions for the superconducting gap were imposed to solve the DBdG equations as the superconducting gap is fixed at a finite value in the superconducting region and abruptly reduces to zero in the graphene region. However, this does not take into account the reduction of the superconducting gap by inverse-proximity effects near the graphene–superconductor interface, or depairing by injected bias current through the superconductors. Tight-binding DBdG equations were solved self-consistently at a zero temperature limit [18,19], and later for a broader temperature range [20], predicting more realistic GJJ behavior in terms of parameters such as superconducting phase, junction length, temperature, Fermi energy, and pairing symmetries.

Another framework for considering GJJs was introduced by González and Perfetto [19,21]. In this scheme, Josephson coupling is considered as the propagation of Cooper pairs through graphene over a long distance ($L \gg W$). This approach is complementary to conventional descriptions of Josephson coupling in terms of Andreev reflections at the interfaces and it is more convenient to discuss many-body effects such as interactions and thermal effects. A crossover temperature $T^* \sim v_F/k_B L$ was predicted, above which the propagation of Cooper pairs is disturbed by thermal fluctuations and marks the onset of power-law decay of the Josephson current. Also, the oscillatory behavior of the Josephson current as a function of junction length was predicted. A similar approach was applied to investigate the different behavior of GJJs based on monolayer and bilayer graphene [22]. In the bilayer case, the Josephson current contribution from the A- and B-sublattices of graphene showed a strong oscillatory behavior, which was ascribed to the interference between conduction- and valance-band electrons. A perturbative Green's functions method in the framework of the path integral was used to calculate the Josephson current with a tunneling Hamiltonian between the superconductor and the graphene [23]. The temperature dependence and junction length dependence for short to long junction limits were examined. By separating the tunneling Hamiltonian into intravalley and intervalley scattering terms, the Josephson current contribution from each scattering type was independently studied. The presence of two valleys leads to an oscillation of Josephson current as a function of junction length. However, this prediction relies on the assumption of atomically sharp interfaces between the superconductor and the graphene, and the period of the spatial oscillation is predicted to be on the atomic scale. Thus, its experimental realization is very challenging and requires atomic-scale controls on junction dimensions.

To date, there have been few theoretical studies on how GJJs are affected by various disorders in graphene. The difficulty to do this arises from the broken translational symmetry by disorders. Recently, a new method of solving self-consistent BdG equations by expanding the Nambu Gor'kov Green's functions in terms of Chebyshev polynomials (Chebyshev-BdG method [24]) was adopted. This method was able to compute the Josephson current and MARs of GJJ by imposing realistic disorders in graphene, such as atomic vacancies, ripples, or charged impurities [25]. While different types of disorder suppress the Josephson current and smear the MAR features, short-range scatterers such as atomic vacancies or resonant impurities are the most influential since they induce intervalley scattering, which destroys the interference of time-reversed electron-hole pairs.

Another type of disorder is the ripple in graphene of inhomogeneous strain, which deforms the distance and the hopping amplitude between carbon atoms and generates a pseudomagnetic field. In contrast to a real magnetic field, a pseudomagnetic field points towards opposite directions for different valleys, so it is considered to break the effective time reversal symmetry defined within a single valley, but not the real time reversal symmetry. Therefore, a pseudomagnetic field destroys the interference between time reversed electron-hole pairs and suppresses Josephson currents. Owing to the very large mechanical strength and flexibility in graphene [26], deformations up to 25% are possible [27] and Landau levels formed by the strain-induced pseudomagnetic fields have been experimentally observed in graphene nanobubbles [28], opening new pathways to study strained graphene. Although no direct experimental observations on the interplay between superconductivity and strained graphene have been made so far, several theoretical predictions have been discussed [29-33].

The last scattering mechanism involves charged impurities, which are modeled as spatially varying chemical potentials known as electron/hole charge puddles. Suppression of Josephson current was shown to be more pronounced for shorter ranges of charge puddles, which enhances the intervalley scattering. The combination of spatially random pseudomagnetic fields generated by ripples was suggested as a dominant dephasing mechanism to completely suppress Josephson current in GJJs near the Dirac point [34]. The existence of ripples was also independently confirmed by the observation of the weak anti-localization (WAL) effect in the normal state.

Here we discuss the experimental works done on GJJs at different length regimes. First, we focus on diffusive GJJs, which are fabricated with either exfoliated graphene on a silicon oxide ($SiO_2$) substrate or using CVD-grown graphene. Later, the ballistic GJJ fabricated with the high-mobility graphene encapsulated between hexagonal boron nitride (BN) crystalline flakes is discussed.

Early works usually focused on rather short diffusive regimes [9,15,16,34-36] as the smaller superconducting gap of Al or Ta meant that $\xi_s$ was longer than the junction length. However, longer diffusive regimes have been more accessible recently through the use of larger gap superconductors, such as lead (Pb) or niobium (Nb) [17,37-40]. Dubos *et al.* [41] calculated the full length dependence on Josephson current for a diffusive JJ made from a non-relativistic normal metal. A universal curve for the

$I_cR_N$ product on $E_{Th}/\Delta$ from short ($E_{Th}/\Delta > 1$) to long ($E_{Th}/\Delta < 1$) regimes was calculated using Usadel's theory with ideal interfaces (Figure 5a). However, in practice, the universality can be easily broken by nonideal interfaces because of sample-specific interfacial potential barriers from disorder, a Fermi level mismatch, or a Fermi velocity mismatch will dramatically affect the Josephson coupling.

Unlike metallic proximity JJs, GJJs can be tuned by gate voltage *in-situ*, which changes the carrier density, Fermi wavelength, mean free path and thus $E_{Th}$ without altering the interfacial properties. Ke *et al.* [40] exploited the gate tunability of graphene to vary $E_{Th}$ by about two orders of magnitude. The measured values of $I_cR_N$ in Pb-based GJJs for three different junction lengths all scaled linearly with $E_{Th}$, as expected in the long diffusive regime. However, the ratio of $I_cR_N/E_{Th}$ was found to be 0.1~0.2, which is much smaller than the theoretically predicted value of 10.82. The reduced value of $I_cR_N$ was attributed to the enhancement of the electron dwell time in the junction due to a low interface transmission coefficient, the origin of which is unknown. A full range of Josephson couplings from short to long diffusive regimes were studied by Li [42] by accessing a ratio of $E_{Th}/\Delta$ over three orders of magnitude, from 0.01 to 10. This was done by combining the effects of tuning the gate voltage to change $E_{Th}$ *in-situ* and using the superconducting materials Al, Nb, and tungsten-rhenium alloy (ReW) to tune $\Delta$ *ex-situ*. Surprisingly, all the $I_cR_N$ values from seven different devices fall into a single universal curve (Figure 5b). Although the absolute value of $I_cR_N$ deviated from theoretical prediction, two asymptotic behaviors qualitatively correspond to the short and long junction limits. $I_cR_N$ linearly scales with $E_{Th}$ in the long-junction regime ($E_{Th}/\Delta < 1$), while it becomes saturated in the short-junction regime ($E_{Th}/\Delta > 1$). The qualitative disagreement of the value of $I_cR_N$ between experiment and theory, with a reduction factor of ~ 10, is again attributed to the imperfect interface. However, the detailed mechanisms that cause this are not clear. Hammer *et al.* [43] predicted that the value of $I_cR_N$ would be sensitively affected by *r*, the ratio of the contact resistance to the graphene channel resistance. However, this prediction contradicts the observed universal behavior of $I_cR_N$, even for values of *r* varied by a factor of 50 using a backgate. The Josephson current is significantly suppressed even for an interface barrier with a small resistance (or, small *r*). Although a numerical calculation with tight-binding BdG Hamiltonian showed an improved agreement with the experiment, it seems obvious that a better understanding is required.

For ballistic GJJs, Andreev bound states (ABSs) are formed in graphene, leading to Josephson coupling. In the short junction limit, only one pair of ABSs exist within $\Delta$, whereas for the long junction limit, there appears to be multiple ABSs with an energy level spacing of $\pi\hbar v_F/L$, resulting in an exponential scaling with temperature, $I_c(T) \propto \exp(-k_BT/\delta E)$ with $\delta E \approx \hbar v_F/2\pi L$ [44]. After significant improvements of graphene quality by adopting the encapsulation technique, Borzenets *et al.* [45] systemically studied ballistic GJJ from short to long junction limits. They fabricated multiple GJJs, the length of which varied from 200 nm to 2 μm, crossing $\xi_s$ (~ 550 nm) with $\Delta$ = 1.2 meV of molybdenum rhenium alloy (MoRe). For the shortest junction ($L$ = 200 nm), the temperature dependence of the Josephson current, $I_c(T)$, followed a convex-shaped dependence, as expected in the short-ballistic limit. For longer junctions, they extracted the energy scale $\delta E$ from the exponential temperature dependence of Josephson current for six different GJJs, and showed that indeed $\delta E$ scaled with $1/L$ but not with carrier density or mobility (Figure 5c). The experimental observation of scaling behavior with temperature agrees with the ballistic theory qualitatively. However, there is a quantitative disagreement of about an order of magnitude in terms of the Josephson current, despite the ballistic nature of graphene and the normal-state transmission coefficient at the interface reaching 0.9. This is a similar behavior to the diffusive GJJs introduced above, in that the measured Josephson current is more suppressed in comparison to theoretical expectation. A better explanation for this discrepancy requires more realistic models for GJJs. More experimental evidence on the nature of Josephson coupling, such as the current-phase relationship, is discussed in the next section. These efforts are important to investigate the peculiar behavior of GJJs and also to improve the Josephson coupling for applications in superconducting hybrid quantum devices.

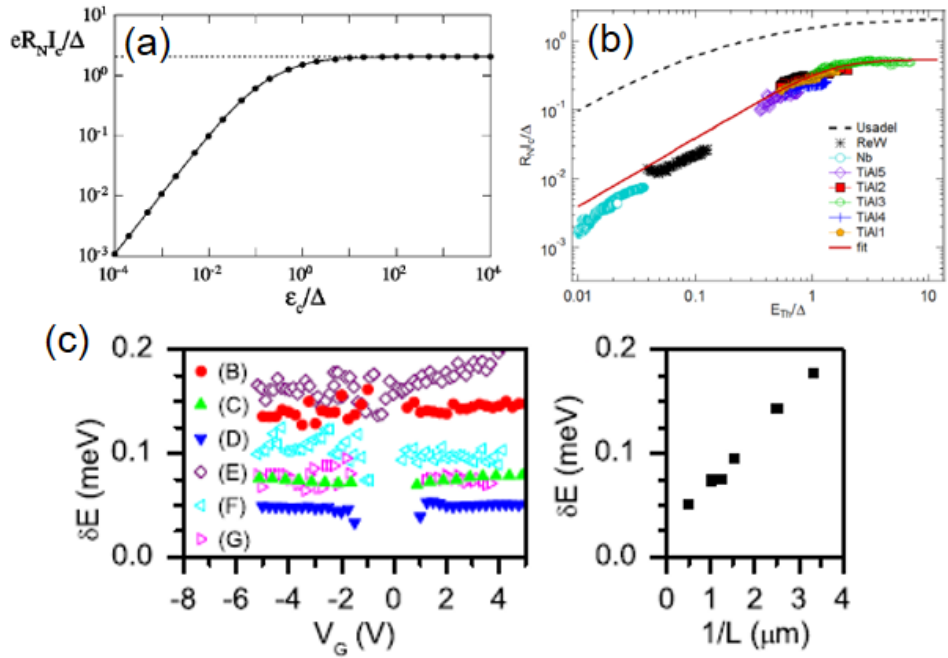

Figure 5. (a) Theoretical calculation by Dubos *et al.* [41]. (b) The relationship between $I_cR_N/\Delta$ and $E_{Th}/\Delta$ for seven diffusive graphene-based Josephson junctions (GJJs) of different junction lengths and different superconducting materials [42]. The dotted line is identical to the curve in (a). (c) (Left panel) The energy

scale δ$E$ as a function of back-gate voltages extracted from the ballistic Josephson junctions with different length $L$. (Right panel) The energy scale δ$E$ shows linear dependence with $1/L$ [45].

## 2. Fabrication

### 2.1. Mechanically exfoliated graphene on amorphous substrates

Soon after thin graphene layers were extracted from graphite using mechanical exfoliation techniques in 2004 [1,46], the first GJJ was fabricated with exfoliated monolayer graphene on a SiO$_2$ substrate with aluminum (Al) as a superconducting material [9,47] (Figure 4). Owing to its low melting temperature (~ 700°C) compared to most other superconductors, Al is preferred for making delicate superconducting contacts with less internal stress and less radiation heating during evaporation from the molten source. Nonetheless, first-principles calculations predicted that Al would weakly bind to graphene at a relatively large distance (~ 3.4 Å) [48] and result in an electrically bad contact, which indeed was experimentally demonstrated with epitaxially grown graphene on SiC(0001) [49]. Therefore, a few-nm-thick titanium (Ti) adhesion layer was inserted between the graphene and Al. This adhesion later played an important role to obtain electrically transparent contacts and realize Josephson coupling across the graphene layer. Although Ti has been widely used as an adhesion material for realizing GJJs [9-18], other adhesion materials were also used, such as platinum (Pt) [16], palladium (Pd) [35,37-40,42,45,50], and tantalum (Ta) [51]. Sometimes an adhesion layer was not used [17,45,52,53]. Although Al has good properties for electrical contacts, it has a low superconducting critical temperature ($T_c$ ~ 1 K), low superconducting gap ($\Delta_s$ ~ 0.1 meV), and low critical magnetic field ($H_{c2}$ ~ 0.1 T). This limits the accessibility of Al to investigate the unique properties predicted in superconducting hybrid structures of graphene and the opportunity for applications, as it often demands the milliKelvin range of temperatures. Therefore, various superconducting materials have been investigated to achieve GJJs at higher operating temperatures or at higher magnetic fields. Pb [17,37,38,40,50], with $T_c$ ~ 7 K and $H_{c2}$ ~ 0.2 T, is one example of a superconducting material. However, extra care should be taken since Pb is easily oxidized in solvents or by moisture in ambient conditions. Also, Pb tends to get thinner with suppressed superconductivity as it is more mobile on graphene than on SiO$_2$ substrates during thermal deposition [17]. Such diffusive behavior is common for materials with low melting temperatures. This problem can be avoided either by quenching the evaporated material with precooled substrates [54] or by putting adhesion layers such as Ti down before the deposition of Pb. Sometimes a few per cent weight fraction of indium (In) can be mixed with Pb for a better morphology [17,55]. Tantalum (Ta) of $T_c$ ~ 2.5 K and $H_{c2}$ ~ 2 T [16], Nb [39,42] of $T_c$ ~ 8 K and $H_{c2}$ ~ 4 T [56], and ReW of $T_c$ ~ 6 K and $H_{c2}$ ~ 8 T [39,42] were also demonstrated to exhibit Josephson coupling across graphene. For materials with high melting temperatures (~ 3000°C), radio frequency/direct current (RF/DC) sputtering methods are usually preferred because thermal evaporation can be very violent and possibly damage the graphene and resist polymers by radiative heating.

Although monolayer graphene is expected to have high mobility due to the suppression of back-scattering, most GJJs made from graphene exfoliated onto SiO$_2$ substrates showed a diffusive behavior. It has been suspected that charged impurities trapped between graphene and SiO$_2$ and polymer residues left on graphene during the fabrication processes create electron-hole puddles, smearing the Dirac point and reducing the mobility and mean free path [57-59]. Du *et al*. [15] analyzed that the magnitude and the backgate dependence of the $I_cR_n$ product of GJJs on SiO$_2$ are well explained by the diffusive model with a mean free path (~ 25 nm) estimated from the normal-state conductance. There are two approaches to avoid impurities between graphene and the substrate. One is to remove the substrate by suspending graphene. Another is to replace amorphous SiO$_2$ substrate with atomically flat and dangling bond-free substrates such as BN. Each of these methods will be discussed in Sections 3.2.2 and 3.2.3, respectively. It is also discussed that GJJs with a smaller lead separation give a shorter mean free path, so it is likely that the quality of graphene is degraded by the deposition of metals. Furthermore, it is understood that the work function difference between graphene and the metal causes an uncontrollable doping of graphene near the metal contacts [48,60,61]. Premature superconducting-to-normal switching induced by thermal and electromagnetic noise is also considered, which is discussed in detail in Section 3.4. To avoid polymer residues on graphene, Shailos *et al*. [62] demonstrated lithography-free superconducting contacts for graphene exfoliated on an aluminum-oxide substrate. Tungsten (W) wires of $T_c$ ~ 4.3 K were directly deposited on few-layer graphene (less than seven layers) by decomposing metallo-organic vapor (tungsten hexacarbonil) with a Ga focused ion beam. Although clear Josephson coupling was not fully established, a zero-bias differential resistance dip and MAR behavior showed the existence of the proximity effect on few-layer graphene. The failure to observe the Josephson coupling was attributed to interfacial contamination of insulating amorphous carbon clusters doped with Ga, which are deposited during the focused-ion-beam growth process.

2.2. Suspended graphene

Trapped charged impurities between graphene and the substrate are not the only source of impurities, interfacial phonons [63] and corrugation induced by a rough surface of the substrate [64-66] are also factors which substantially limit the mobility of graphene. Therefore, suspending graphene by eliminating substrate was suggested and experimentally demonstrated to enhance the carrier mobility of graphene (Figure 6a) [67-69]. This can be done by immersing the whole graphene structure with attached electrodes in buffered oxide etchant containing hydrogen fluoride (HF). The capillary action quickly draws the etchant to the interface between the graphene layer and the $SiO_2$ substrate, while the electrodes blocked the etching underneath. Therefore, the etching time is carefully tuned so that the graphene can be fully suspended before the electrode structure detaches from the substrate. One of the drawbacks of using HF is that not only does it etch $SiO_2$, but it also etches certain metallic materials, including Ti and Al. To solve this problem, Ti can be replaced by chromium (Cr) as an adhesive layer [67,69], or the electrodes can be protected by a polymethyl methacrylate (PMMA) polymer layer from exposure to the HF [68,70]. Tombros *et al*. [71,72] developed a different method of suspending graphene without using HF, which is especially helpful for fabricating suspended GJJ, as many superconducting materials are not compatible with HF processing (Figure 6b). Polydimethylglutarimide (PMGI)-based organic polymer coated onto a solid substrate is first used as a substrate for the graphene. This polymer has strong chemical resistance to a wide range of solvents so that standard electron beam lithography and electrode lift-off processes can be performed. However, PMGI is still sensitive to high doses of electron beam or deep-UV radiation so that the polymer underneath the graphene layer can be selectively removed. Another method involving two different polymers, PMMA and copolymer methyl methacrylate (MMA), was introduced and used to fabricate suspended GJJs (Figure 6c) [10,73]. The graphene is exfoliated onto a PMMA spacer layer that it covers a $SiO_2$/Si substrate, and the MMA is used to define the electrodes. To make a smooth connection from the suspended graphene to the electrodes on the $SiO_2$/Si substrate, the electron beam exposure dose was controlled considering the electron beam sensitivities of PMMA and MMA. To avoid a collapse of the suspended graphene due to the surface tension of the solvents during the drying process, either critical point drying techniques [67], or low surface tension solutions such as isopropyl alcohol or hexane can be used [10,68,71,72]. Since the mechanical stability becomes more serious with longer graphene devices, GJJ with lengths smaller than 1 μm are considered to be relatively safe from structural collapse. The quality of graphene immediately after suspension is usually comparable to graphene supported by $SiO_2$ substrates due to polymer residues left on the graphene layer during fabrication. Therefore, post cleaning of the residues is crucial for suspended GJJs. Joule heating by using large amounts of current (density of 0.3 ~ 1.0 mA/μm) sent through the device can raise the temperature of the graphene up to a few thousands of Kelvin [74,75] and subsequently removes surface contaminants. The metallic contacts act as heat sinks so that the temperature near the contacts is lower than that of the central region. Therefore, contaminants can pile up near metallic contacts, which make it hard to evaporate these contaminants without overheating the central region, resulting in a burning out of the graphene layer. This significantly lowers the yield of preparing high-mobility suspended graphene devices.

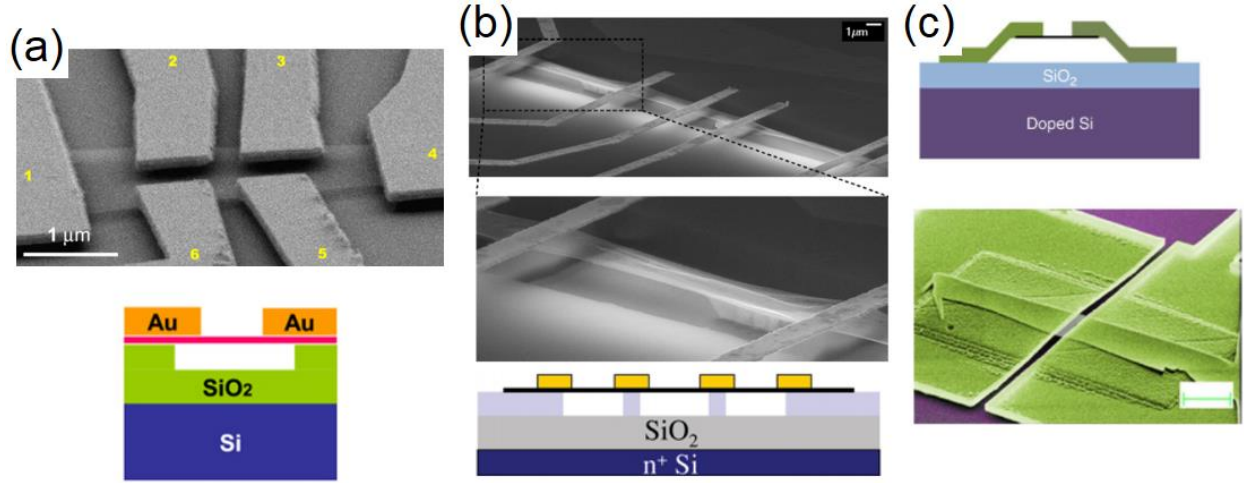

Figure 6. (a) Scanning electron microscope (SEM) image (upper panel) and schematics (lower panel) of a suspended graphene device made using hydrogen fluoride [67]. (b) SEM image (upper panel) and schematics (lower panel) of graphene device made using a polymer sacrificial layer [71]. (c) Schematics (upper panel) and SEM image (lower panel) of a suspended graphene JJ [73].

2.3. Graphene on hexagonal boron nitride substrates

Another intriguing idea to enhance the mobility of graphene is to replace the $SiO_2$ substrate with a BN crystal. BN is an attractive dielectric substrate compared to conventional amorphous $SiO_2$ in many ways. It can be easily exfoliated due to van der Waals stacking and makes an ideal substrate for graphene with a large (~ 6 eV) bandgap, and a small (1.7%) lattice mismatch to graphene. Moreover, its atomically flat surface suppresses corrugation of graphene and the inert nature of the surface minimizes possible charge traps caused by absorbed impurities. Furthermore, the surface optical phonon energy of BN is almost twice as large in comparison to $SiO_2$, which suggests improved mobility at high temperatures. However, the uncontrollable position of exfoliation of BN and graphene causes technological challenges. Various methods to transfer graphene into designated positions on BN have been developed. The first generation of G/BN van der Waals heterostructures were fabricated by transferring graphene onto a prepared BN flake on a $SiO_2$ substrate with the use of sticky polymer materials [76,77]. While there are variations in the details of the materials and setups, the basic principle is as follows. Graphene is exfoliated onto a thin polymer layer, which is placed on a transparent glass slide. After facing the graphene side down, it is still possible to locate the graphene and the polymer layer through the glass slide, so that graphene can be aligned to the BN flake using optical microscopy and micromanipulators. After placing the graphene down onto the BN flake, the polymer layer is removed by solvents and then by either annealing above 300°C in flowing $H_2$/Ar or by current annealing. This post annealing process is particularly important to achieve high graphene mobility by removing the polymer residue on the graphene. However, the annealing process reduces the yield of high quality graphene and may degrade the delicate superconducting contacts, which could be a reason why there have been so few reports on GJJ with G/BN heterostructures.

In 2013, noticeable improvements to the fabrication of high quality graphene were made by replacing the polymer-sticking layers with BN flakes (Figure 7) [78]. Instead of using polymer to pick up the graphene for transfer, a (top) BN flake can be used to pick up graphene easily due to van der Waals interactions between the flat surfaces of the two materials. The BN/G structure is placed onto another (bottom) BN flake, resulting in a stack of BN/G/BN. The main advantage of this method is that graphene is never exposed to any polymers, touching only clean BN flakes during the transfer process. As graphene is fully encapsulated by BN flakes, it is free from contamination during the subsequent lithography processing and does not require annealing, significantly enhancing the fabrication yield and mobility of graphene. Electrical contacts can be made onto a one-dimensional (1D) edge of the graphene, which is exposed by etching across the stack of BN/G/BN with a slanted profile [78]. The first reported method for 1D side contacts involved a separate lithography process for defining electrodes after the etching process. Hence, the graphene edges were contaminated by resist polymers and hindered the formation of transparent electrical contacts. Later, a number of studies reported improved methods to form the contacts, where electrodes are deposited onto freshly exposed graphene edges immediately after the etching process. Here, the lithographically defined etch mask also serves as a lift-off resist layer for defining the electrodes. JJs fabricated with graphene prepared on $SiO_2$ substrates are usually of a diffusive regime, where the mean free path ($l_{mfp}$) is shorter than the junction length ($L$), limiting the accessible scope of physics. However, recent developments in nanofabrication techniques such as suspension of graphene or BN encapsulation have enabled demonstrations of GJJ in the ballistic regime ($l_{mfp} > L$) with assorted superconducting materials such as thermally evaporated Al [70,79] or sputtered MoRe [45,52,53], Nb [51,80], or NbN [81], depending on the physical phenomenon being studied. The improved quality of graphene can reveal unique fundamental physics that can bridge superconductivity with relativity. Furthermore, these techniques provide powerful tools to realize quantum devices with superconductor-graphene heterostructures.

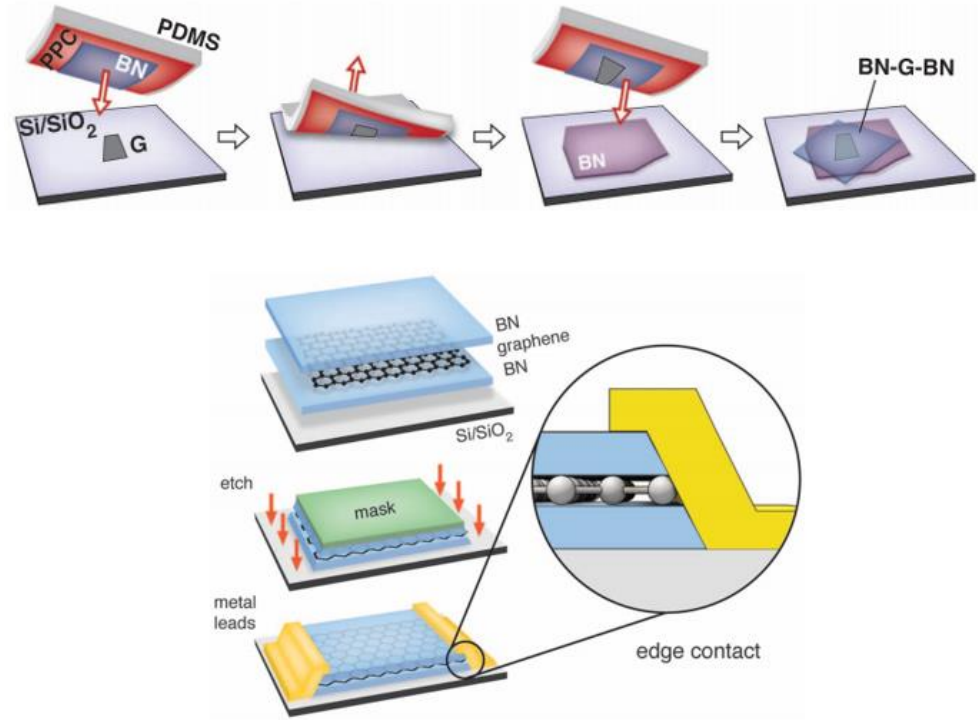

Figure 7. (Upper panel) Dry transfer sequence for the encapsulation of graphene in hexagonal boron nitride layers. (Lower panel) Schematics for making one dimensional contacts to the exposed edge of graphene [78].

3. Current-phase relation (CPR)

From the microscopic point of view, proximity Josephson coupling is established by successive Andreev reflections of quasi-particles between two superconducting interfaces, as depicted in Figure 8. Quasi-particles undergoing Andreev reflections gain the information of the macroscopic quantum phase of each superconductor, resulting in the ABS, $E_A(\phi)$, which has a dependence on the phase difference ($\phi$) between two superconductors. ABS determines the Josephson current flowing through the junction as $I_J(\phi) = (2e/\hbar)(\partial E_A/\partial \phi)$ and is known as the current-phase relationship (CPR), the measurement of which can reveal unique information about the microscopic processes which influence Josephson coupling.

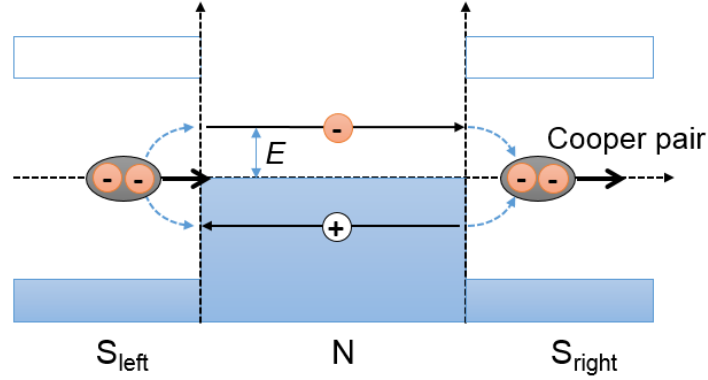

Figure 8. Microscopic description of Josephson coupling. A Cooper pair transfers from the left superconductor ($S_{left}$) to the right ($S_{right}$) via a normal metal (N) by successive Andreev reflections.

Although the simplest case of tunneling JJs show sinusoidal CPR, deviations from this sinusoidal relationship have been observed in systems with various types of conducting spacers, including JJs with highly transparent atomic contacts [82-85], metals [86], nanowires [87], or 2D electron gas (2DEGs). Moreover, the periodicity of the CPR can be different from $2\pi$ for exotic topological JJs. In this section, we introduce two different theoretical approaches to CPR and the experimental efforts to measure the CPR.

The first theoretical approach was to solve the DBdG equations with rigid boundary conditions. This method was used for a short-junction regime in the zero-temperature limit by Titov and Beenakker [5], and later for more general conditions of the junction length and arbitrary temperature by Hagymasi *et al*. [88]. Figure 9a summarizes the CPR as a function of different parameters such as length, temperature, and doping level. Only for the special case at the Dirac point and zero temperature, there exists an analytic form for the CPR ($\phi_c = 0.63\pi$) given by [5],

$$I_J(\phi) = \frac{e\Delta}{\hbar} \frac{2W}{\pi L} \cos(\phi/2) \tanh^{-1}[\sin(\phi/2)], \quad \text{(Equation 1)}$$

for the wide junction limit ($L \ll W$). Interestingly, this CPR calculated for ballistic GJJs is formally identical to that of a disordered normal-metal JJ for $k_F l_{mfp} \rightarrow 1$. This correspondence is another manifestation of the relativistic behavior of graphene and is termed pseudo-diffusion. For general parameters, however, the CPR cannot be expressed analytically in terms of $\phi$, and should be calculated numerically. As shown in Figure 9a, in general, the CPR at low temperatures is positively skewed ($S > 0$, $\phi_c > \pi/2$) away from the conventional sinusoidal CPR ($S = 0$, $\phi_c = \pi/2$), and with increasing temperature near the critical temperature the CPR approaches the harmonic form. Here, the skewness is defined as $S = 2\phi_c/\pi - 1$, with the critical phase, $\phi_c$, at which Josephson current become maximum. Away from the Dirac point, the longer junction shows a more skewed and linear CPR, which resembles the sawtoothed CPR ($S = 1$), predicted for a long ballistic normal-metal JJ. Hagymasi *et al.* also found that for small $I_c$, the skewness, $S$, grows linearly with $I_c$ while the temperature decreases. This serves as a convenient comparison between theory and experiment since $I_c$ can be controlled *in-situ* by gate voltages in GJJs. The CPR has been measured using phase-sensitive SQUID interferometry techniques for short diffusive GJJs [89], and later for short and quasi-ballistic GJJs and long diffusive GJJs [90]. Indeed, their skewness of the CPR showed that there was a clear linear relationship with $I_c$, regardless of $I_c$ being tuned by either temperature or by gate voltage. However, it is not clear to what extent the calculation for ballistic GJJs applies to experimental cases of quasi-ballistic or diffusive GJJs.

Another theoretical approach is based on the self-consistent tight-binding BdG (TB-BdG) formalism for more realistic modeling which takes into account the inverse proximity effect [18] and depairing by current [20], both of which decrease the superconductivity of superconducting electrodes. The CPR calculated by the TB-BdG method in different regimes is shown in Figure 9b. This approach is not available for large Josephson currents due to the phase drop in the superconductors, as shown in the lower panels of Figure 9b. When the Josephson current is small (near the Dirac point, at high temperatures, or in a long junction regime), the TB-BdG method does not show a notable change from the DBdG method. The needs for the TB-BdG method appear to be more significant for shorter junctions and at lower temperatures, where the Josephson current is big and depairing plays an important role. Indeed, the TB-BdG method explains the experimental observation of skewness of the CPR better than the DBdG method [90]. The TB-BdG theory predicts that for highly doped GJJs, the skewness changes even its sign to negative ($S < 0$, $\phi_c < \pi/2$, negatively skewed), nonetheless this has not yet been verified experimentally.

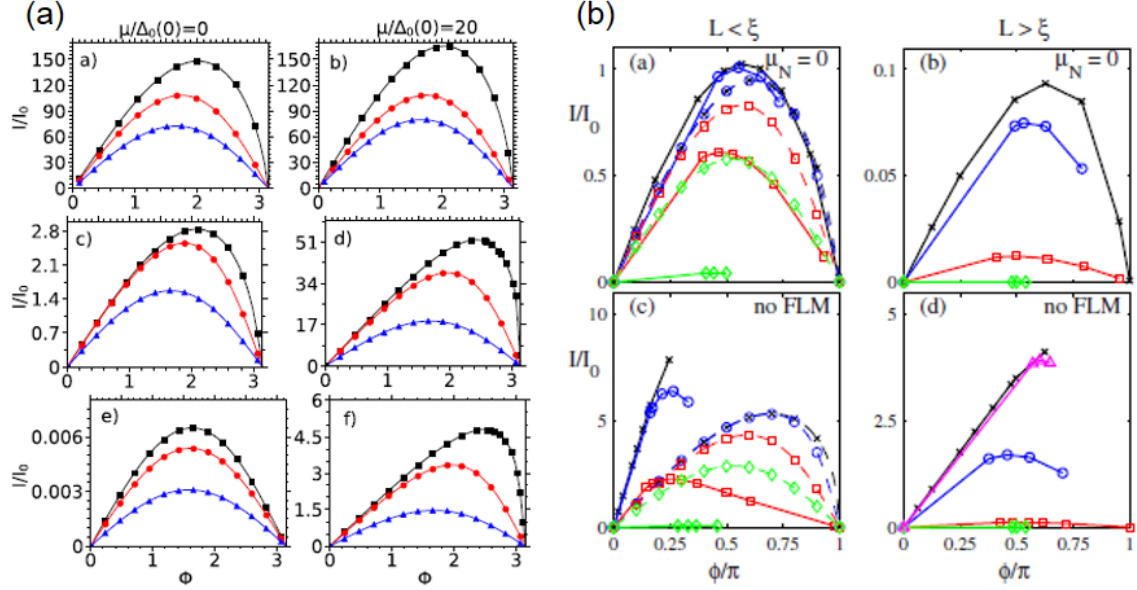

Figure 9. (a) Current-phase relation (CPR) calculated by the Dirac-Bogoliubov-de Gennes (DBdG) method [88]. (b) CPR calculated by the self-consistent tight-binding BdG [20].

The magnitude of the Josephson current can be measured easily by biasing current and monitoring output voltages, whereas CPR measurements should involve a phase-sensitive scheme. Some information about the CPR can be indirectly obtained from the Fraunhofer patterns of $I_c$, Shapiro steps, or $I_c$-modulation of a DC-SQUID (superconducting quantum interference device). The first experimental attempt to gain information on the CPR was done using a DC-SQUID, made of two GJJs connected in parallel via superconductors (Figure 10a) [35,91]. The modulation of $I_c$ as a function of the external magnetic field $B$ contains indirect information of the CPR at each junction. A skewed modulation of $I_c(B)$ was observed, however, this cannot be considered conclusive evidence for the skewness of the CPR. This is because the self-inductance of the physical superconducting loop ($L \sim 15$ pH) is too big to explain the observed skewness without assuming a skewed CPR. Moreover, two unknown CPRs for each GJJ determine the $I_c(B)$ of the DC-SQUID, making it hard to extract the CPR from the observed $I_c$ modulation. Such ambiguity can be removed by adopting an asymmetric DC-SQUID with minimum self-inductance, which consists of a GJJ under investigation and another reference tunneling JJ with a sinusoidal CPR (Figure 10b). When the Josephson coupling of the tunneling JJ is much larger than that of the GJJ, the $I_c$ of the DC-SQUID is dominated by the tunneling JJ and the phase difference across the tunneling JJ approaches $\pi/2$ near $I_c$. Then, the phase difference across the GJJ is uniquely determined by the condition of total phase accumulation around the SQUID ring to be $2\pi n$, with integer values of $n$. This technique has been utilized to observe a highly skewed CPR of mechanical break atomic JJs made of aluminum [83,84]. A similar method was also adopted for a vertical GJJ which consisted of monolayer graphene sandwiched between two aluminum superconducting electrodes [92]. In this vertical structure, where the graphene is not gate tunable, the junction length becomes atomically short and easily meets the condition for the short-ballistic regime. The measurement of a highly skewed CPR was interpreted as evidence for the short-ballistic nature of the vertical GJJ, along with the measured $I_cR_N$ product (~ 2.6$\Delta$/e) being larger than the theoretical limit of the short-diffusive regime (~ 2.1$\Delta$/e). Another kind of CPR measurement involves an RF-SQUID and consists of one GJJ in a superconducting ring [89,90]. An RF-SQUID is coupled to an extremely sensitive magneto-sensor (DC-SQUID), as shown in the schematics of Figure 10c, so that the phase across the GJJ can be inferred by the signal from the DC-SQUID sensor.

Since the fabrication of ballistic GJJs has become more accessible, the CPR for ballistic GJJs can be investigated using DC-SQUID interferometry composed of two GJJs, which are made with MoRe superconductors and graphene encapsulated by BN [93]. An asymmetric DC-SQUID was fabricated by tuning individual GJJs with local top gates for the CPR measurement on marginally short ($L \leq \xi_s$) and ballistic ($L \ll l_{\text{mfp}}$) GJJs. The skewness $S$ reached up to ~ 0.3 in the electron ($n$)-doped regime, whereas $S$ oscillated around the value of 0.2 with a carrier density in the hole ($p$)-doped regime. The magnitude of the skewness and the oscillatory behavior of the $p$-doped regime were consistent with the TB-BdG numerical calculation when the formation of the $n$-$p$-$n$ cavity in the graphene is taken into account. Owing to the work function difference of graphene and superconducting metals, graphene near the superconducting contacts is inevitably $n$-doped and hard to control by a backgate. When the backgate voltage is applied for $p$-doping, only the central part of the graphene is tuned to be $p$-doped, and causes the formation of an $n$-$p$-$n$ cavity in the graphene. Such phase-coherence phenomena, termed Fabry-Perot interference, are discussed in more detail in Section 4.2. Lastly, it is worthwhile to note that the SQUID loop should be maintained non-hysteretic (the screening parameter $\beta_L = 2\pi L I_c/\Phi_0 \leq 1$) for correct CPR measurements. Here, $\Phi_0 = h/2e$ is the magnetic flux quantum. Such conditions can be achieved either by minimizing the self-inductance or by minimizing $I_c$. At the same time, however, $I_c$ should be large enough to avoid thermal smearing effects for the condition of Josephson coupling energy $E_J = \hbar I_c/2e \gg k_B T$.

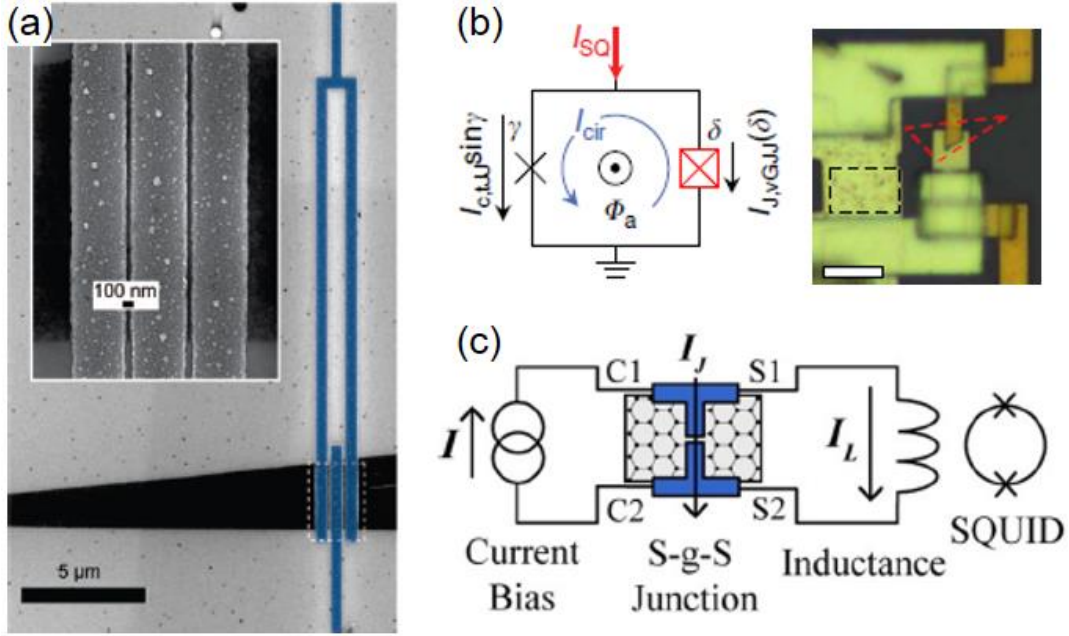

Figure 10. (a) SEM image of a DC-SQUID made from two GJJs in a superconducting ring (blue) [35]. (b) (Left panel) DC-SQUID made from one vertical GJJ under investigation (red box with a cross, red dotted triangle in the right panel) and a tunneling JJ (black cross, black dotted rectangle in the right panel) in a superconducting ring [92]. (Right panel) Optical image of the device. (c) RF-SQUID made from a single GJJ and a superconducting inductive coil inductively coupled to the DC-SQUID magneto-sensor [90].

4. Switching dynamics of the phase particle

With the macroscopic quantum nature that is inherited from superconductivity, a JJ is a special system where the microscopic order parameter behavior is directly controlled by the macroscopic parameters, such as voltage and current. Specifically, JJs are typically modeled as a parallel electrical circuit (Figure 11a) made up of capacitance, $C_J$, between two superconductors, resistance, $R_J \sim R_N$, through the barrier, and non-linear inductance, $L_J = (\hbar/2e)/(I_c\cos\phi)$, which represents the Josephson coupling assuming a sinusoidal CPR. In this resistively and capacitively shunted junction (RCSJ) model, the dynamic behavior of JJs are described by a fictitious phase particle in a so-called washboard potential $U(\phi)$, with a dissipation that scales with $1/R_J$ (Figure 11b). The coordinate and mass of the phase particle correspond to $\phi$ and $m = (\Phi_0/2\pi)^2 C_J$, respectively. A biasing current tilts the washboard potential so that eventually the trapped phase particle can escape the potential barrier, $\Delta U$, and start to move down the potential. Some literature refers to this as a phase slip. The voltage, *V*, across the junction is proportional to the 'speed' of the phase particle according to the AC Josephson relationship, $V = (\hbar/2e)(\partial\phi/\partial t)$ with time *t*. Therefore, the dynamics of the phase particle can be detected by measuring the voltage; a zero voltage indicates the trapped state and finite voltages indicate the phase-particle running state. Owing to quantum and thermal fluctuations, or external noise, the switching current $I_s$ represents the escape of the phase particle and displays a stochastic behavior. Also, $I_s$ is always smaller than the fluctuation-free critical current $I_{c0}$, at which $\Delta U$ vanishes. The ratio of the small-oscillation frequency (or plasma frequency), $\omega_{p0} = (2eI_{c0}/\hbar C_J)^{1/2}$, to the energy dissipation rate, $1/R_J C_J$, defines the quality factor, $Q = \omega_{p0} R_J C_J$, of the JJ and parameterizes the degree of dissipation. Experimentally, underdamped JJs ($Q \gg 1$) show hysteretic current-voltage characteristics; where the retrapping current, $I_r$, at which the running phase particle gets retrapped in the washboard potential is smaller than $I_s$ due to the large inertia. Overdamped JJs ($Q \ll 1$) display a non-hysteretic behavior; where the phase particle dissipates its energy so quickly that it undergoes successive retrapping at the potential local minimum.

Premature switching induced by thermal fluctuations and/or external electromagnetic noise in GJJs was taken into account by several authors [15,17,36,73] to explain the reduction of $I_s$ and backgate dependence on the $I_s R_N$ product. The mean reduction in switching current can be estimated [94] as $<\Delta I_s> \sim I_{c0}[(E_{fl}/2E_J)\ln(\omega_p \Delta t/2\pi)]^{2/3}$ in the smaller fluctuation limit $E_{J0} \gg E_{fl}$. Here, $E_{J0} = \hbar I_{c0}/2e$ is the Josephson coupling energy, $E_{fl}$ is a characteristic fluctuation energy, $\Delta t$ is the time spent sweeping the bias current through the dense part of the distribution of $I_s$, and $\omega_{p0} = (2eI_{c0}/\hbar C_J)^{1/2}$ is the plasma frequency of the JJ. To take into account external noise, $E_{fl}$ was written as $E_{fl} = k_B(T + T_{EM})$, where $T_{EM}$ is an effective temperature increase due to electromagnetic noise [15,73]. The logarithmic term is weakly sensitive to the parameters, yielding ~ 21–28, with typical parameters of $\Delta t \sim$ 0.1–100 s and $\omega_{p0} \sim 10^{11}$–$10^{12}$ Hz. Owing to the large logarithmic term, $I_s$ experiences a major reduction when $E_{fl}/E_{J0} \sim 0.1$. This is because the reduction in $I_s$ becomes more significant near the Dirac point, where $E_{fl}/E_{J0}$ is larger, and the fluctuations result in a much stronger backgate dependence on $I_s$ than theoretical expectation [5].

The measurement of escaping statistics of the metastable state gives important information about thermal and quantum fluctuations. This problem was firstly discussed in tunneling JJs by Fulton and Dunkleberger [95], where they analyzed switching current distribution $P(I_s)$ measured by linearly increasing the bias current at a rate of $dI/dt$. The expression for the escaping rate is given as

$$P(I) = \Gamma(I)(dI/dt)^{-1}\left[1 - \int_0^I P(I)dI\right], \qquad \text{(Equation 2)}$$

where $\Gamma(I)$ is the phase particle. The escape processes driven by quantum and thermal fluctuations are termed macroscopic quantum tunneling (MQT) and thermal activation (TA), respectively. The escaping rate for TA ($\Gamma_{TA}$) [95] and MQT ($\Gamma_{MQT}$) to the lowest order in $1/Q$ [96] are given by

$$\Gamma_{TA} = a_t \left( \omega_p / 2\pi \right) \exp(\Delta U / k_B T),$$
$$\Gamma_{MQT} = 12\omega_p \sqrt{3\Delta U / 2\pi \hbar \omega_p} \exp\left(-7.2(1+0.87/Q)\Delta U / \hbar \omega_p \right),$$
(Equation 3)

where $a_t = (1+1/4Q^2)^{1/2} - 1/2Q$ is a damping-dependent factor of the order unity, $\omega_p = \omega_{p0}(1-\gamma^2)^{1/4}$, and $\Delta U = 2E_{J0}[(1-\gamma^2)^{1/2} - \gamma \arccos\gamma]$ with normalized bias current $\gamma = I/I_{c0}$. Here, a conventional sinusoidal CPR is assumed for $\omega_p$ and $\Delta U$ of a GJJ, which is justified by the fact that the CPR of a diffusive GJJ is close to sinusoidal shape. More detailed analysis conducted by Lambert *et al.* [97] showed that $\omega_p$ and $\Delta U$ of the short ballistic JJ with skewed CPR are essentially similar to the results from a conventional tunneling JJ.

As $\Gamma_{TA}$ is exponentially suppressed with decreasing $T$, while $\Gamma_{MQT}$ remains almost constant, a crossover between MQT and TA appears around the temperature $T^*_{MQT} \sim a_t \hbar \omega_p / 2\pi k_B$, below which $\Gamma_{MQT}$ dominates over $\Gamma_{TA}$ and the total escaping rate saturates. The temperature dependence on the standard deviation $\sigma$ of the switching current distribution usually follows a power law $\sigma \propto T^{2/3}$ in the TA regime [95] and saturates in the MQT regime below $T^*_{MQT}$. Lee *et al.* [98] studied the temperature and gate-voltage dependences on $P(I_s)$ of the GJJ and showed the MQT of the phase particle in the proximity of the JJ was similar to conventional tunneling JJs [99]. Moreover, *in-situ* electrical control on $T^*_{MQT}$, or equivalently $\omega_p$, was demonstrated following the predicted relationship of $T^*_{MQT} \propto \omega_p \propto I_{c0}^{1/2}$ by exploiting the gate tunability of graphene. Microwave spectroscopy measurements also confirmed the existence of quantized energy levels ($\hbar\omega_p$) formed in the metastable state of the washboard potential and their tunability with the backgate. These results opened up the possibility of using GJJs as gate-tunable superconducting phase quantum bits. Later, full control of the escaping mechanism with backgate was also demonstrated at a fixed temperature by forming a *p-n* heterojunction in the GJJ by using a global backgate and a local topgate [34].

Such macroscopic quantum coherence behavior is only possible for underdamped JJs ($Q > 1$). However, $C_J$ with high $Q$ values are not well defined in proximity JJs, as opposed to the case of parallel plate capacitance in a tunneling JJ. A possible origin of the capacitance in a proximity JJ was suggested by Angers *et al.* [100] and suggested that the correlated electron-hole pair relaxation time is no longer set by $R_J C_J$ time, but by $\hbar/E_{Th}$ (or $\hbar/\Delta$) in a diffusive (or ballistic) transport regime. A correspondence is then achieved by replacing $C_J$ with an 'effective' junction capacitance $C_{J,\text{eff}} = \hbar/R_J E_{\text{Th}}$. Lee *et al.* showed that $C_{J,\text{eff}}$ inferred from $T^*_{MQT}$ increased linearly with $1/R_N$ as the value of $\omega_p$ was tuned by the backgate, which supports the effective capacitance scenario. There is another explanation for the hysteretic current-voltage characteristics in GJJs with $Q \sim 4I_c/\pi I_r$ higher than 1. The capacitance may come from the capacitive coupling between the source and drain pads via a conducting backgate electrode made from highly doped silicon [15,73]. The hysteretic current-voltage characteristics routinely observed in proximity JJs can also originate from Joule heating of the junction above $I_s$ and the subsequent reduction of $I_r$ [101]. However, the sharp transition from a zero voltage state to a finite voltage state as well as the insensitive $I_s/I_r$ ratio with different magnitude of $I_s$ [100] or junction area [37] were suggested as evidence for the intrinsic nature of the hysteresis in JJs. So far, the origin of junction capacitance in GJJs has not been systematically investigated or clearly understood.

Coskun *et al.* [38] focused on the thermal activation mechanism in underdamped GJJs ($Q \approx 4$) through the measurement of $P(I_s)$. It was shown that $\sigma$ scaled anomalously with temperature as $\sigma \propto T^\alpha$, with $\alpha$ as low as 1/3, in contrast to the usual case of $\alpha = 2/3$ [95] of tunneling JJs and intrinsic JJs of high-$T_c$ superconductors [102]. This anomalous behavior was attributed to the temperature dependence of $I_s$, which is usually neglected within the temperature range for studying TA in tunneling JJs. The authors also generalized Kurkijarvi's theory [103] for the scaling of $\sigma$ for arbitrarily smooth CPR and found the power law of $\sigma \propto T^{2/3} I_c^{1/3}$. Taking into account the reduced standard deviation, $\sigma/I_c^{1/3} \propto T^\alpha$, the exponent $\alpha$ was found to be near 2/3 independent of backgate. Interestingly, as $I_c$ can be controlled by backgating, the exponential relationship between $\sigma$ and $I_c$, $\sigma \propto I_c^{1/3}$ was directly confirmed at fixed $T$. As the Josephson coupling decreases at higher temperatures, the dissipation can play a significant role in the dynamics. Even for underdamped JJs of $E_J \gg k_B T$, if the phase particle has escaped at a bias current smaller than $I_m \sim (4/\pi Q')I_c$ [104], it would not have enough energy left after the dissipation to overcome the next potential maxima and thus diffuse from one well to another. Here, $Q'$ is the quality factor at the plasma frequency. Since the phase particle gets retrapped below $I_m$, the junction does not give a measurable voltage. Such processes are termed underdamped phase diffusion (UPD), and are signified by a sudden drop of $\sigma$ at $T^*_{TA} \sim E_{J0}(1-4/\pi Q)^{3/2}/30 k_B$ [98]. $T^*_{TA}$ at different backgate voltages also followed an expected linear relationship of $T^*_{TA} \propto E_{J0} \propto I_{c0}$ as $Q$ was almost constant (5 ~ 6) for the entire backgating range. At even higher temperatures, where $k_B T \sim E_{J0}$, the phase particle can diffuse even without a current bias via overdamped phase diffusion. This results in a finite zero-current resistance $R_0$, even at bias currents below $I_c$. Borzenets *et al.* [37] used GJJs made of Pb with a higher critical temperature to investigate the overdamped phase diffusion at relatively higher temperatures. The temperature dependence of $R_0$ was confirmed to follow the theoretical prediction of $R_0(T) \propto T^{-1}\exp(-2E_{J0}/k_BT)$ [105]. The analysis of $R_0$ on $E_{J0}$ at a fixed temperature, enabled by backgating, gave unique information to directly identify underdamping behavior at low frequencies and overdamping behavior at the plasma frequency.

Recently, a comprehensive theoretical work has been conducted on the dynamics of phase particles in short ballistic GJJs considering thermal and correlated fluctuations, such as Gaussian or colored noise [106].

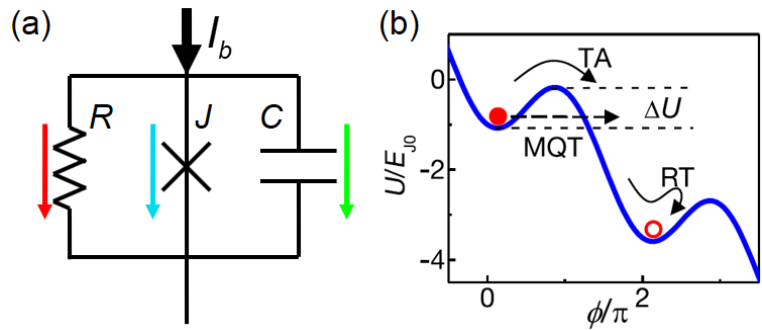

Figure 11. (a) Schematics of the representative electrical circuit for a resistively and capacitively shunted junction (RCSJ) model. (b) The phase particle (red ball) initially trapped in the local minimum of the washboard potential of the RCSJ model. It can escape the potential barrier $\Delta U$ by either macroscopic quantum tunneling (MQT) or thermal activation (TA). The escaped phase particle can be retrapped (RT) if it dissipates enough energy at the following local minimum [98].

## 4. Mesoscopic physics

In this section, we focus on the phase-coherent behavior of the superconductor–graphene interface. The proximity effect of hybrid structures, made by contacting various materials to superconductors, has been studied intensively with the continuous development of nano-fabrication techniques and as the understanding of the coherence effect and topological properties in mesoscopic systems are enhanced. This is especially interesting for the unique properties of graphene as it presents an opportunity to explore the interplay between superconductivity and relativistic quantum mechanics.

1. Andreev reflections in graphene

The proximity effect is the generic term for phenomena appearing at superconductor/normal (S/N) interfaces and is often used to describe the superconducting-like behavior of the normal metal near the superconductor. From a microscopic point of view, an incident electron from N with energy less than $\Delta$ may be reflected at the SN interface as a hole, while a net charge of $2e$ enters the superconductor as a Cooper pair. A converted hole moves in the opposite direction following the path of the incident electron, as shown in Figure 12 (retro-reflection), because the hole moves in the opposite direction to its momentum while the momentum along the interface is conserved. This process, known as Andreev reflection [107], converts a normal dissipative current into a dissipationless supercurrent and creates a correlation between the incident electron and the reflected hole in the normal metal. From a macroscopic point of view, this is responsible for the superconducting-like behavior of normal metals near the interface. However, the retroreflection of a hole is only an approximation for typical metals where the Fermi energy $E_F$ (a few eV) is much bigger than the typical superconducting gap $\Delta$ (a few meV).

Unusual Andreev reflections were predicted at the superconducting contacts with monolayer [108] and bilayer graphene [109] where the $E_F$ of graphene can be smaller than $\Delta$ near the Dirac point. In this regime ($E_F < \Delta$), an incident electron in the conduction band is converted into a hole in the valance band (interband). Since the converted hole in the valance band moves in the same direction as its momentum, the group velocity along the interface is unchanged and results in a specular Andreev reflection (Figure 12). One of the striking features of monolayer graphene is the unit Andreev reflection probability for perpendicular incident electrons, even with a large Fermi level mismatch at the junction. This is a consequence of the chirality conservation. An Andreev-converted hole conserves its original chirality by residing in the same sublattice, while a reflected electron without the Andreev process requires either intravalley scattering from one sublattice to the other or intervalley scattering with a large momentum transfer. The same explanation can be applied to the perfect transmission of normally incident electrons through the potential barrier in graphene, which is called Klein tunneling. The correspondence between Andreev reflection and Klein tunneling in graphene is extensively discussed by Beenakker *et al*. [110,111].

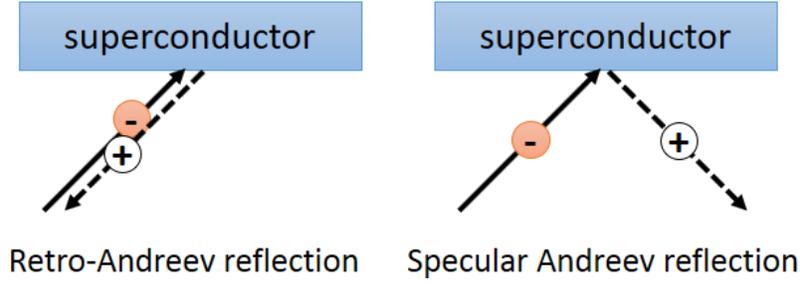

Figure 12. Retro-Andreev reflection (left panel) and specular Andreev reflection (right panel).

The major difficulty with the experimental observation of specular Andreev reflections is graphene Fermi energy fluctuations ($\delta E_F$) larger than $\Delta$. For example, graphene prepared on a $SiO_2$ substrate has a carrier density fluctuation of $\delta n \sim 10^{11}$ $cm^{-2}$, which corresponds to $\delta E_F \sim 50$ meV [112,113]. Recent development in the fabrication of suspended graphene [67,68] or BN-supported graphene devices [76,78] significantly reduced the fluctuations to $\delta n \sim 10^9$ $cm^{-2}$, which corresponds to $\delta E_F \sim 5$ meV [113]. The signature of specular Andreev reflections was first observed in a bilayer graphene system [114]. The superconducting contact was made by transferring mechanically cleaved $NbSe_2$ ($\Delta \sim 1.2$ meV) onto graphene and current annealing was performed to achieve a low contact resistance. Bilayer graphene was encapsulated with BN to achieve a small energy fluctuation of $\delta E_F < \Delta$. Owing to its large density of states near the Dirac point, bilayer graphene has much smaller $\delta E_F$ than monolayer graphene at a given value of $\delta n$. When backgating is used to tune $E_F$ to be smaller than $\Delta$, there can be three distinctive regimes depending on the energy of an incident electron ($eV_{ns}$): i) for $eV_{ns} < E_F$, the reflected hole resides in the conduction band (retro Andreev reflection); ii) for $eV_{ns} > E_F$, the reflected hole resides in the valance band (specular Andreev reflection); or iii) for $eV_{ns} = E_F$, the possible state in which the reflected hole resides vanishes, resulting in suppressed conductance (Figure 13a). Dip features in conductance near the Dirac point were predicted in Ref. [109] and later rigorously calculated in Ref. [115] for bilayer graphene, and similarly for monolayer graphene in Ref. [108]. As shown in Figure 13b and c, the conductance dip follows the condition of $eV_{ns} = E_F$, separating the two regimes of retro and specular Andreev reflection. However, the direct observation of the specularity of reflected holes is still lacking, partly because the ballistic motion of electrons would be largely hindered by charged impurities near the Dirac point.

Monolayer graphene has the advantage of a relatively simple linear band structure, whereas bilayer graphene may have a gap-opening near the Dirac point and multiple discrete low energy pockets due to the trigonal warping effect near the Dirac point [116]. Furthermore, the chirality conservation of graphene ensures a perfect Andreev reflection for perpendicularly impinging electrons. Sahu *et al*. [117] investigated the monolayer graphene system with NbSe$_2$ superconducting contacts. However, the larger fluctuation ($\delta E_F$ ~ 15–35 meV) estimated from the residual carrier density made it harder to satisfy the condition for specular Andreev reflection. Although conductance suppression near the Dirac point was demonstrated, this cannot be definitive evidence of specular Andreev reflections as they failed to show the revival of the conductance at lower bias voltages due to large energy fluctuations.

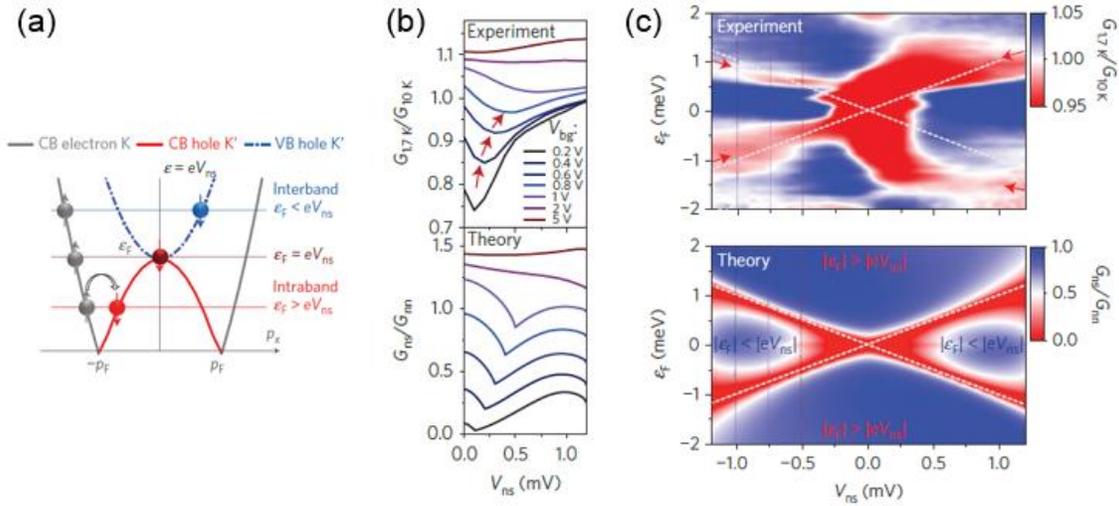

Figure 13. (a) Schematic of the Andreev reflection. (b) Normalized differential conductance with the value in the normal state. The upper panel shows the experimental values and the lower panel shows the theoretical calculation [114].

There have been several suggested experimental setups to distinguish or control specular Andreev reflections, although no experimental realizations have been made to date. The first example is the thermal conductance measurement over a long GJJ [118]. Similar to the quantum confinement in a narrow conductor sandwiched between reflective barriers, a normal metal confined by two superconducting barriers forms ABSs (Figure 14a). For an $E_F$ of graphene smaller than $\Delta$, a series of specular Andreev reflections form Andreev modes in a long junction limit ($E_{Th} \ll \Delta$), which are a superposition of electrons in the conduction band and holes in the valance band (Figure 14b). The Andreev modes are charge-neutral, so they cannot transport charge along the interface. Instead, they can convey energy, so the thermal conductance can be finite with a strong dependence on the phase difference $\phi$. The maximum thermal conductance appears for $\phi = \pi$, reaching one-half of the thermal quantum conductance, $\pi k_B^2 T/6\hbar$, per spin and valley degrees of freedom. A similar propagating Andreev mode was suggested to realize pure spin current by injecting spin-polarized electric current [119]. In this setup, the current carries no charge but only spin. The second example is the four-terminal geometry graphene/superconductor device [120] depicted in Figure 14c, where retro and specular Andreev reflections can be controlled by the phase difference $\phi$ of two superconducting electrodes (terminals 2 and 4). When incoming electrons from terminal 1 (blue arrows) undergo a retro Andreev reflection, two reflected holes (broken black arrows) accumulate $\phi$, exhibiting a destructive interference for $\phi = \pi$ with a suppressed retro Andreev reflection. Whereas for specular Andreev reflections, destructive interference happens at $\phi = 0$ with suppression of the specular Andreev reflection. This is due to the extra phase $\pi$, which originates from the fact that the incident electron and reflected hole are in different energy bands. For larger graphene where the diffraction effect vanishes, reflected holes go to either terminal 1 or 3 depending on the specific kinds of Andreev reflection, providing an experimental method to detect the nature of Andreev reflections. Later, Cheng *et al*. [121] performed numerical analysis on the effects of electron-hole inhomogeneity and showed that a specific kind of Andreev reflection can be distinguished even for large energy fluctuations of $\delta E_F/\Delta \sim 100$. Lastly, Aharonov-Bohm interference in graphene rings with superconducting contacts (Figure 14d) has been discussed to differentiate the types of Andreev reflection [122,123]. The fundamental period of the interference for usual metals is a flux quantum $h/e$, whereas it is half of the value ($h/2e$) for retro Andreev reflections due to the Cooper pair charge $2e$. For a specular Andreev reflection, however, the reflected hole follows a different path with accumulating opposite Aharonov-Bohm phase resulting in the fundamental period of $h/e$, according to the numerical analysis up to the first order. This behavior distinguishing the different types of Andreev reflection is robust against bulk disorder due to the topological nature: the periodicity is determined depending on the arm through which the reflected hole traverses, so that impurities cannot alter this behavior unless strong scattering occurs across different arms.

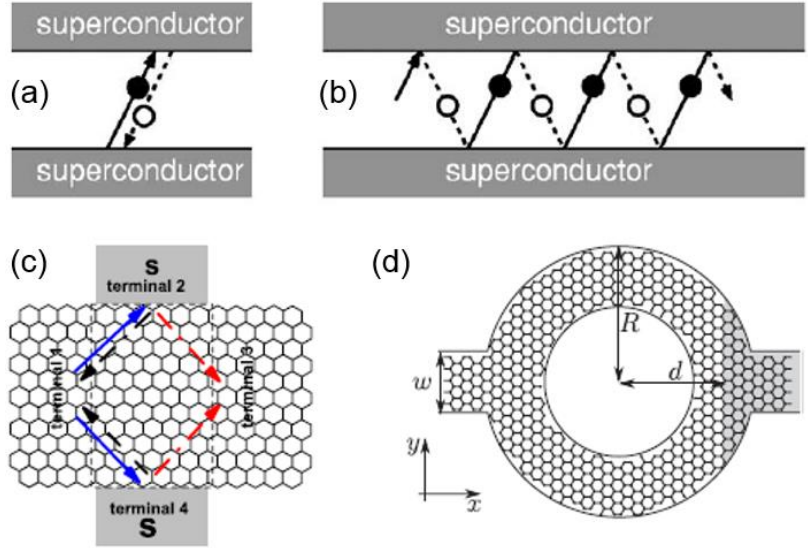

Figure 14. (a) Andreev bound state forms with electron (filled circle) and converted hole (open circle) by retro Andreev reflection. (b) A propagating Andreev mode is formed by specular Andreev reflections when the Fermi energy is smaller than the superconducting gap [118]. (c) A schematic of the four-terminal graphene-based superconducting hybrid system. The blue arrows represent the incident electrons while the black (red) broken arrows represent reflected holes with retro (specular) Andreev reflections [120]. (d) A graphene-ring interferometer schematic with a superconducting contact (shaded region) [122].

Andreev bound states formed in proximitized graphene coupled to superconductors were directly measured using tunneling probes. Dirks *et al.* [61] made tunneling Pb superconducting contacts onto graphene through an aluminum oxide tunneling barrier, as shown in Figure 15a. Pb superconducting electrodes not only induced proximity superconductivity in the graphene, but locally doped the graphene underneath by the work function difference of graphene and Pb. The combination of the superconducting proximity effect and the quantum confinement by local doping resulted in the formation of ABS. For usual tunneling superconducting contacts to a normal metal, the differential conductance vanishes below the superconducting gap because of the absence of a quasiparticle density of states. However when ABSs are formed inside of the superconducting gap, the resonant tunneling though them results in the subgap conductance features shown in Figure 15b. As the backgate voltage shifts the energy levels of the quantum confinement, the subgap ABS features progressively change accordingly. Another tunneling measurement was made onto a graphene JJ embedded in a superconducting ring (Figure 15c) [124]. In contrast to the previous experiment, local doping of the graphene was avoided by using graphite electrodes, which have a similar work function to graphene. In this geometry, the phase sensitive information of the ABS was directly accessed by controlling the superconducting phase difference across the GJJ via an external magnetic field (Figure 15d). By introducing the novel concept of supercurrent spectral density, Bretheau *et al.* demonstrated the conversion from the experimental information on density of states of graphene to the CPR of JJs. Such a method provided a new way to measure the CPR of JJs by combining Andreev physics with the Josephson effect.

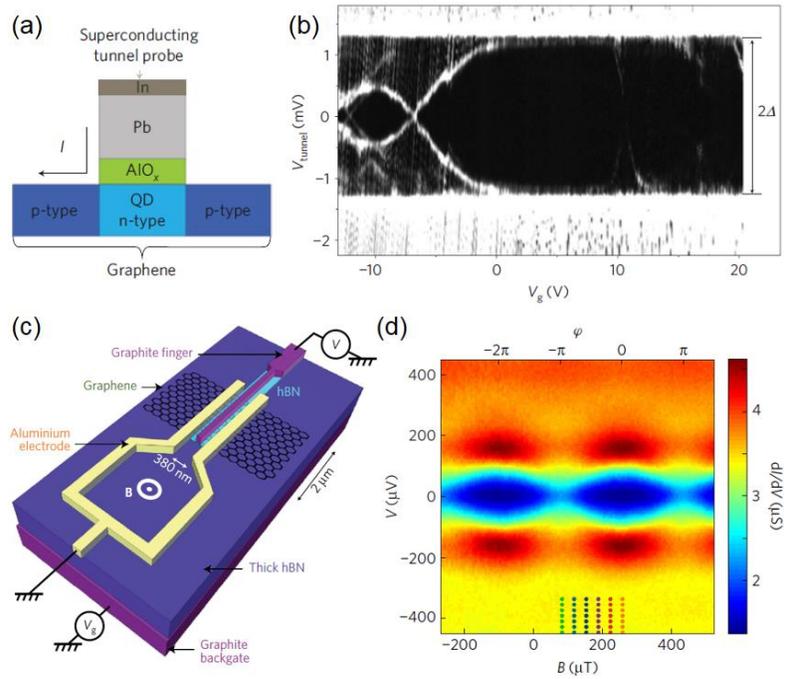

Figure 15. (a) Tunneling lead (Pb) superconducting onto the graphene. (b) Differential conductance as a function of the backgate voltages ($V_g$) [61]. (c) A graphene-based RF-SQUID device with a tunneling probe made from graphite and thin boron nitride as a tunneling barrier. (d) Differential conductance as a function of external magnetic field $B$ [124].

## 2. Phase-coherent phenomena

In mesoscopic conductors, electrons can preserve their quantum mechanical phase information while they travel over a certain length scale (phase coherence length $L_\phi$) and exhibit phase coherent phenomena, which originates from their wave nature. Particularly, superconductor–graphene heterostructures are remarkable systems to understand the phase-coherence behavior of Andreev pairs of induced superconductivity with tunable transport parameters. Interference of Andreev quasiparticles in graphene-NbTiN junctions [125] manifested a zero-bias conductance enhancement exceeding a factor of two, which is the limit of Blonder-Tinkham-Klapwijk (BTK) theory. This was explained by 'reflectionless tunneling', where the transmission across the junction barrier is enhanced by the diffusive nature of the normal metal [126]. As shown in Figure 16a, the electron $e_1$ which fails to reflect as a hole $h_1$ can have a second chance of Andreev reflection into $h_2$ due to scattering from impurities in the graphene. Constructive interference of $h_1$ and $h_2$ increases the probability of an Andreev reflection of $e_1$. It was also demonstrated that such conductance enhancement by interference could be suppressed by a finite voltage corresponding to the phase coherence length or an external magnetic field corresponding to the flux quantum threading the loop of the diffusive path.

In conventional superconductors at low temperature, $L_\phi$ usually exceeds the superconducting coherence length and limits the superconducting correlations. On the other hand, graphene as a highly tunable conductor provides an opportunity to study the interplay between superconductivity and dephasing mechanisms. Specifically, graphene Andreev interferometers (Figure 16b) have been studied to elucidate the roles of various mesoscopic length scales. Electrons at the point P can propagate either to the left or right superconductor and reflect as Andreev holes, which later come back to the point P following the same trajectories of the electrons in a time-reversed way. The conductance of the system is enhanced (suppressed) when the reflected holes show constructive (destructive) interference. Such interference depends on the difference in phase that the holes acquire from each superconductor and can be controlled by the external magnetic flux threading a big superconducting loop. Deon *et al.* [127] investigated graphene Andreev interferometers (Figure 16c and d) focusing on the temperature, bias voltage, and backgate dependences on the reentrance of the superconducting proximity effect [128-130]. This counterintuitive behavior of non-monotonic dependence on the Andreev interference amplitude originates from two competing origins. One is conductance enhancement as more Cooper pairs propagate into the N with increasing superconducting coherence length. Another is a conductance decrease as the quasi-particle density of states decreases in N as the pair potential increases when the superconducting coherence length exceeds the size N. Compared to the normal metals used in previous conventional devices, graphene gives a significant advantage for experimentally investigating the reentrance effect with continuously tunable transport parameters such as $L_\phi$. When the Fermi level is far away from the Dirac point (where $L_\phi$ is presumably much larger than the distance between superconductors $L$), the normalized amplitude of the Andreev interference stays essentially the same as predicted by theory and the onset energy of the reentrance effect then matches with the Thouless energy of the graphene. On the other hand, when the Fermi level approaches the Dirac point, the normalized amplitude decreases and the onset energy of the reentrance effect then shifts to a larger value. The authors attributed the suppression of amplitude as a result of $L_\phi$ being shortened below $L$. The progressive shift of onset energy was explained by the fact that the penetration of Andreev pairs is now limited by $L_\phi$ and the relevant energy scale becomes $\hbar D/L_\phi^2$. The linearized Usadel equations qualitatively explained both dependencies of interference amplitude and the onset energy. The reentrance effect of a graphene system as an $n$-$p$-$n$ junction contacted to superconductors has been theoretically discussed in terms of the interplay between Klein tunneling at the $n$-$p$ interface and Andreev reflection at the S/N interface [131]. Kim *et al.* [132] also performed transport measurements on a similar graphene Andreev interferometer (Figure 16c) in a different regime, where $L_\phi$ did not play an important role as $L_\phi \gg L$ for the entire range of temperature and backgate voltages. They focused on the locality of the proximity-induced superconductivity by tuning the thermal coherence length, $\xi_T = \sqrt{\hbar D/2\pi k_B T}$. The locality of the proximity effect was originally discussed in mesoscopic JJs of $L_\phi \gg L,W$, where $W$ is the width of JJ [133]. As the aspect ratio $L/W$ increases, the Fraunhofer pattern showed a periodicity of $h/e$ rather than the conventional value of $h/2e$. This anomalous behavior was explained by the nonlocal Josephson coupling from different positions along the superconducting interface [134]. In a graphene Andreev interferometer, the Fraunhofer pattern appears as an envelope of the faster Andreev interference oscillations. As $\xi_T$ of the graphene was tuned by varying the backgate voltage, the shape and periodicity of the Fraunhofer pattern changes between local or nonlocal forms. When $\xi_T \sim L$, the interference of two Andreev processes happening at the same $x$ position along the superconducting interface has a major contribution, resulting in a local Fraunhofer pattern. However, for $\xi_T < L$, any pair of points along the interfaces equally contribute to the Andreev interference, giving a nonlocal Fraunhofer pattern.

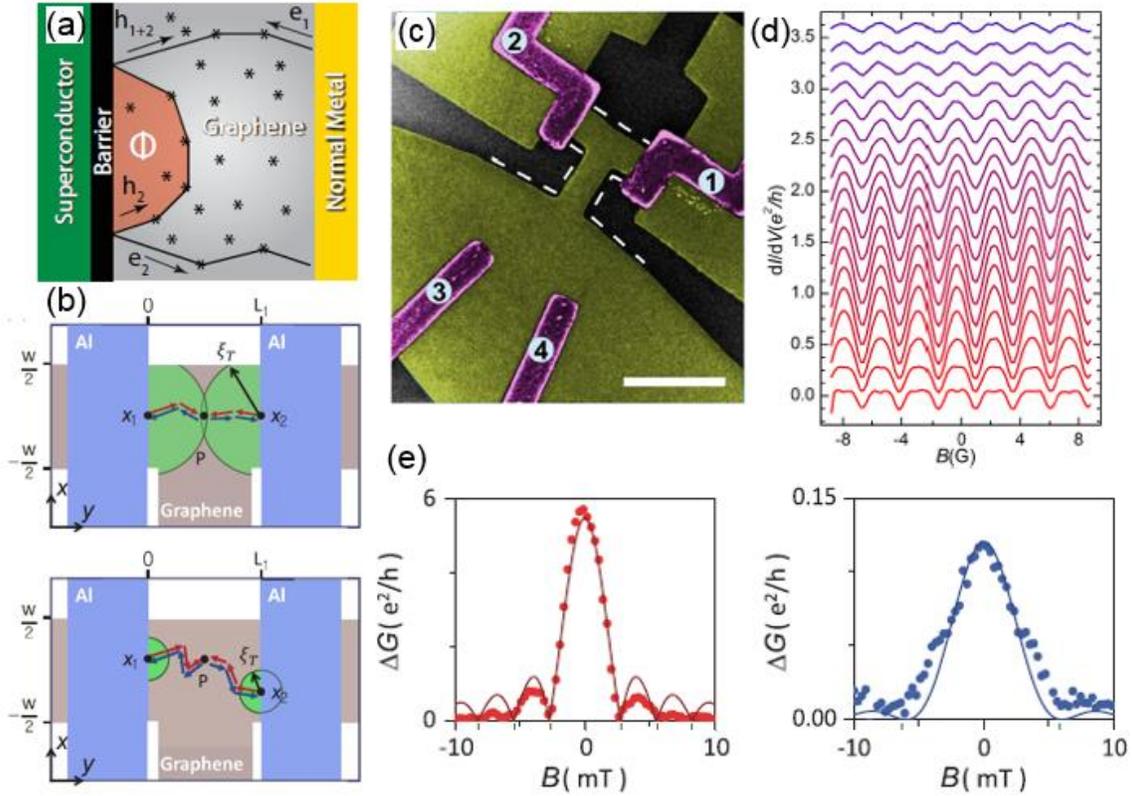

Figure 16. (a) A schematic of reflectionless tunneling [126]. (b) Schematics of graphene Andreev interferometers in local (upper panel) and nonlocal (lower panel) regimes [132]. (c) SEM image of a graphene Andreev interferometer [127]. (d) The conductance oscillation as a function of external magnetic field $B$ for bias voltages from 0 (bottom curve) to 0.5 mV (top curve). Each curve is offset for clarity [127]. (e) Envelope of the conductance oscillation as a function of $B$ for a backgate voltage far away from the Dirac point (left panel, local behavior) and near the Dirac point (right panel, nonlocal behavior) [132].

3. Quantum phase transition

The quantum phase transition, a transition between quantum phases, is only accessed by non-temperature parameters. In contrast to the classical phase transition governed by thermal fluctuations, the quantum phase transition is driven by quantum fluctuations arising from Heisenberg's uncertainty principle. An example quantum phase transition is the superconductor-insulator transition, which has been extensively studied in condensed matter systems for the past thirty years [135]. It is an important subject not only to investigate quantum phase transitions themselves, but also to unveil the mechanisms of emerging high-$T_c$ superconductivity. However, a consensus on its nature has not yet been reached.

At lower dimensions, the quantum fluctuation is enhanced and plays an important role in the system. Therefore, the superconductor-insulator transition has been studied mostly in 2D superconducting thin films. An open question is how the transition evolves with physical variables, such as carrier density mediating the superconductivity and the role of disorder and dissipation in the system. To answer this question, the electric field effect can provide a powerful and versatile method to control the various physical variables without altering the disorder landscape of a single device. However, the field effect is suppressed in typical superconducting thin films due to their large carrier density. In this respect, graphene can provide numerous advantages for studying superconductor-insulator transitions. First, the very low carrier density of graphene enables it to be efficiently controlled by the electric field. In addition, the band gap of bilayer graphene is also controllable with an applied electric field. Second, it is a true 2DEG with exposed surfaces to the environment and chemically inert and stable characteristics. Such properties allow transparent contacts with superconducting materials, which efficiently induces superconducting correlations into the graphene. This is essential to study the superconductor-insulator transition as the graphene is not intrinsically superconducting. Third, the suppressed inverse proximity effect due to the low carrier density and weak intrinsic interactions (such as spin-orbit coupling) would minimize back-action of graphene to the induced superconducting correlations.

Kessler *et al.* [136] introduced a simple and efficient way to make arrays of submicron superconducting islands on exfoliated graphene without complicated lithographic patterning. When tin (Sn) of low melting temperature is deposited on graphene at room temperature, it forms self-assembled islands as shown in Figure 17a due to the poor wettability of graphene. Although the structure of the superconducting clusters is inhomogeneous, this system behaved electrically as a homogenous dirty 2D superconductor. Below the $T_c$ of Sn (~ 3.5 K) the resistance shows a broad transition to the zero-resistance state, which is understood as the Berezinskii-Kosterlitz-Thouless (BKT) transition of a dirty 2D superconductor. As the disorder enhances quantum fluctuations of the order parameter, the system produces antiparallel vortex pairs, the flow of which contributes to the finite resistance even without an external magnetic field. Sn-decorated graphene is a unique system to test the relationship between the BKT transition and the normal resistance in a single device. Below the critical BKT transition temperature, vortex pairs get bound together resulting in zero resistance. The backgate voltage dependence of the BKT transition temperature was shown to follow the theoretical prediction considering the normal-state resistance of the system [137]. Later, Allain *et al.* [138] applied a similar fabrication technique to CVD-grown graphene which has significant disorder. The disorder promotes strong localization of electrons at low temperature, resulting in a superconductor-insulator transition. In contrast to the experiment with exfoliated graphene [136] where the resistance always decreases with temperature, a strong insulating behavior appears near the charge neutrality point as shown in Figure 17b. The separation of superconducting and insulating phases occurs at a sheet resistance of the order of $h/4e^2$, the quantum resistance of the Cooper pair. This critical behavior follows the dirty boson model in a low-dissipation regime. Furthermore, finite-size scaling analysis near the transition shows a universal behavior in the temperature range of 0.6 to 1.2 K, suggesting a quantum phase transition. The exponent $z\nu = 1.18 \pm 0.02$ extracted from the scaling analysis corresponds to the theoretical value of a disordered 2D bosonic system. Here, $\nu$ and $z$ are the correlation length and dynamical critical exponents, respectively, which together characterize the universality class of quantum phase transition. In terms of applications, a dramatic resistance change by more than seven orders of magnitude around the transition is of interest for making sensitive sensors such as transition-edge particle detectors. In addition, the fabrication method is very simple and has been demonstrated on scalable CVD-grown graphene already.

JJ arrays at varying Josephson coupling strengths were predicted to show a quantum phase transition from superconducting to non-superconducting [139-143]. An experimental study [144] with arrays of Nb islands on gold films controlled the proximity Josephson coupling by fabricating multiple devices with differing separation between the superconducting islands. Later, Han *et al.* [145] exploited the tunability of graphene to investigate the influence of quantum phase fluctuations on the superconductivity with a regular JJ array deposited on CVD-grown graphene (Figure 17c). As the backgate gets closer to the Dirac point, the superconducting phase continuously collapses to a weakly insulating phase (Figure 17d). The BKT model correctly describes the temperature dependence of resistance far away from the Dirac point, as in Ref. [136]. However, the resistance starts to deviate from the BKT model as the normal resistance increases to the order of $h/4e^2$ near the Dirac point, where quantum fluctuations become significant. This was attributed to the failure of the BKT model, which considers thermal, but not quantum, fluctuations. Interestingly, increasing quantum fluctuations does not bring the system directly into the insulating phase. An intervening metallic phase appears with the saturating resistance at low temperatures. The zero-bias differential resistance dip indicates the existence of superconducting correlations in this metallic phase. These observations suggest the 2D quantum metal, or so-called Bose metal. In addition, the re-entrance to a superconducting state above the critical field supports the superconducting glass phase [143], resulting from phase frustration between superconducting islands and the mesoscopic fluctuations of the system.

In contrast to monolayer graphene, bilayer graphene can have tunable bandgap by using top and bottom gates. In the presence of electron-hole puddles induced by charge impurities in the supporting substrate, the tunable bandgap of bilayer graphene results in charge puddles of which the separation is tunable, as depicted in Figure 17e. Thus, bilayer graphene can serve as a unique platform to study quantum phase transitions with a varying strength of disorder. For inducing superconducting correlations, bilayer graphene was inserted between two superconducting electrodes forming a JJ [146]. The superconductor-insulator transition can be driven either by tuning the carrier density with a fixed bandgap, as shown in Figure 17f, or by tuning the bandgap with a fixed carrier density. The finite-size scaling analysis of both the temperature and electric field dependence was performed to determine the universality class of the quantum phase transition. The universality class turns out to be temperature dependent; the transition shows a classical percolation of charged bosons above 400 mK, while it shows a quantum percolation below 400 mK. The direct and continuous control of disorder enabled an estimation of the electron temperature at low temperatures, which in turn enabled the identification of the crossover between classical and quantum percolation in a single device. Using graphene as a tunable dissipation source for studying dissipation-driven superconductor-insulator transitions was theoretically suggested by Lutchyn *et al.* [147]. They suggested the use of graphene as a tunable resistor, which is in tunneling contact to the JJ arrays. When the graphene is conductive, it populates gapless quasiparticles and induces more dissipation into the system, suppressing quantum phase fluctuations and driving a quantum phase transition.

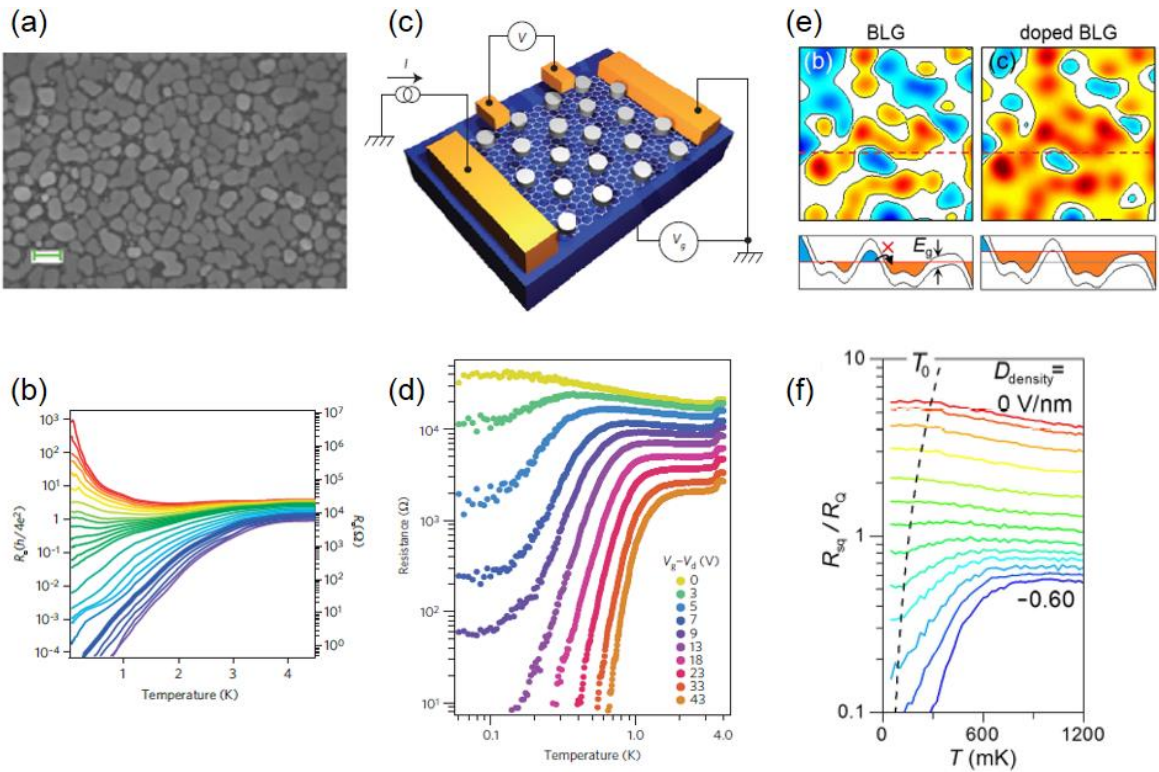

Figure 17. (a) Scanning electron microscopy image of tin-decorated graphene. Scale bar is 300 nm. (b) Temperature dependence of sheet resistance at different backgate voltages from the charge neutrality

point (red) to the high doping regime (blue) [138]. (c) Schematic for a device consisting of triangular superconducting island arrays on graphene. (d) Temperature dependence of the device resistance at different backgate voltages from the charge neutrality point (yellow) to the high doping regime (orange) [145]. (e) Spatially distributed charge puddles of bilayer graphene with a finite bandgap at a nominal charge density of zero (left panel) and a finite value (right panel). (f) Temperature dependence of the square resistance at different carrier densities, from low (red) to high values (blue) [146].

4. Proximity effect in the quantum Hall regime

Two of the most celebrated phenomena in condensed matter physics are the quantum Hall effect and superconductivity, each having been studied extensively for more than 100 and 40 years, respectively. There is a resemblance between these two phenomena, in that they manifest quantum physics at macroscopic scales and they are insensitive to microscopic details. However, their underlying physics are totally different. Curiosity for how the quantum Hall effect and superconductivity will interact has been brought to both theoretical and experimental attention. Theoretical studies have predicted proximity Josephson coupling along the quantum edge states [148], conductance across the superconducting interface on the quantum-Hall edge state with and without disorder [149-151], and the formation of Andreev edge states that propagates along the superconducting interface [152,153]. However, there are a few technical requisitions for experimental studies. The mobility of the 2DEG should be high enough to promote a quantum Hall phase at the magnetic field lower than the $H_{c2}$ of the superconductor. Also, electrical transparency of the superconducting contact should be sufficiently high so that the Andreev process can efficiently happen. Experimental studies only began after nanofabrication technology enabled superconducting hybrid structures of high mobility 2DEGs in GaAs or InAs-based semiconducting heterostructures [154-158]. However, the major obstacle is that a transparent superconducting contact is sometimes tricky to fabricate because of detrimental etching processes, which are necessary for reaching the 2DEG buried below the surface, and the formation of Shottky barriers due to a finite bandgap of the semiconductors. In this regard, graphene is a compelling platform thanks to the exposed nature of the 2DEG and zero bandgap ensuring transparent contacts.

Unlike the 2DEGs of semiconducting heterostructures, the valley degree of freedom of electrons in graphene plays an important role in the Andreev process. The electrons of graphene in different valleys are related by time reversal symmetry, just like electrons consisting of Cooper pairs. Andreev reflections can happen only when incident electrons in upstream edge states and reflected holes in downstream edge states have opposite valley polarization components. Akhmerov and Beenakker [159] theoretically showed that the conductance, $G_{NS}$, across graphene in the $v = 2$ quantum Hall regime and a superconducting contact is determined by the inner product of valley polarizations $v_1$ and $v_2$ of upstream and downstream edge states, as $G_{NS} = 2e^2/h(1-\cos\Theta)$ with $\cos\Theta = v_1 \cdot v_2$. Depending on the crystallographic configuration of the graphene edges, $G_{NS}$ can be $G_{NS} = 4e^2/h$ when both edges are zigzag or both are armchair and are separated by $3n$ hexagons, or $G_{NS} = e^2/h$ when both are armchair separated by $3n + 1$ or $3n + 2$ hexagons. Here, $n$ is an integer number. However, in reality, the intervalley scattering and atomic roughness of the edge randomize the valley polarizations and exponentially suppress the importance of the $\cos\Theta$ term. Full randomization of valley polarization would result in $G_{NS} = 2e^2/h$, which is identical to the conductance with normal contacts.

Rickhaus *et al.* [56] made transparent Nb superconducting contacts onto exfoliated graphene on a $SiO_2$ substrate by optimizing the thickness of the Ti adhesion layer. They focused on the two-probe conductance of the Nb/graphene/Nb device (Figure 18a) in the integer quantum Hall regime. An electron impinging on the superconducting interface can undergo an Andreev reflection by converting into a hole, similar to the case of no magnetic field. When the fermi energy is larger than the chemical potential of the incident electron, the reflected hole resides in the same band as the incident electron. Subsequently the reflected hole has an opposite charge and mass to the incident electron, resulting in a chiral motion the same as the incident electron. As shown in Figure 18b, the two-probe conductance shows the enhancement as the Nb electrodes gradually become superconducting from $B = 4$ T to 3.2 T. In this analysis, the contact resistance of the device was subtracted, such that the two-probe conductance gives the correct quantized values above $H_{c2}$. However, it is not clear how the resistance contribution of the Nb electrodes were treated while the contribution varied above and below $H_{c2}$. Compared to the conductance enhancement of 40% for $v = 6$ and 80% for $v = 10$, $v = 2$ quantum-Hall states showed a relatively small enhancement (10%). This was explained by the strong intervalley scattering. However, at higher filling factors, valley-degenerated edge states support Andreev reflections regardless of the valley polarization and give a large conductance enhancement. For an understanding of the Andreev process in the quantum Hall regime, Komatsu *et al.* [39] focused on differential resistance measurements of a ReW/graphene/ReW device. Although neither a clear supercurrent nor a significant conductance enhancement were observed, the differential resistance showed alternating peak and dip features as the magnetic field or backgate voltages varied. This was attributed to coherent interference of Andreev pairs in diffusive trajectories. The zero-bias differential resistance dip was claimed to be the signature of a Josephson current, but its origin was not undoubtedly clarified.

After the technological advancement of manipulating van der Waals layered materials, investigating the superconducting proximity effect in high quality graphene under high magnetic field has now become possible [51-53,81]. Ballistic JJs were first demonstrated in graphene with superconducting edge contacts of MoRe ($H_{c2}$ ~ 8 T) [52] and Nb ($H_{c2}$ ~ 4 T) [51]. However, neither the MoRe-based or Nb-based GJJs showed any signature of Josephson current or Andreev reflections. This is very surprising since the high mobility of graphene was improved by two orders of magnitude compared to earlier experiments [39,56]. Nonetheless, Shalom *et al.* [51] observed that the Josephson current persists at relatively high magnetic field (~ 0.5 T) where it should have vanished according to the Fraunhofer dependence. The mechanism of this remnant Josephson coupling was suggested to be edge scattering of electrons in cyclotron motion due to the magnetic field. This was supported by the observation of fluctuations in the Josephson current, for which the characteristic energy and magnetic field correspond well to the Thouless energy estimated by the cyclotron radius and the magnetic flux quantum threading the junction, respectively. When the junction gets into the quantum Hall regime, where the cyclotron radius becomes shorter than half of the junction length, no signs of a Josephson current were found. Authors explained that this was because a single chiral edge state cannot support Josephson coupling.

Interest in quantum-Hall–superconducting hybrid systems has been reignited after the acknowledgement of its potential to realize topological superconductivity, a novel quantum phase characterized by non-Abelian anyons. Owing to their exotic braiding statistics, non-Abelian anyons have drawn great interest for their applications in fault-tolerant topological quantum computers. The most experimentally accessible non-Abelian anyons have been Majorana zero modes, the candidates of which include a fractional quantum Hall state [160] of $\nu = 5/2$, intrinsic chiral $p + ip$ superconductors, s-wave superconductors coupled to topological insulators [161], or 1D semiconducting nanowires [162-164]. However, braiding Majorana modes cannot perform arbitrary quantum operations without the help of non-topological quantum operations [165]. This limitation has led to the recent suggestion of utilizing a fractional quantum-Hall/superconducting hybrid system, which is particularly interesting in that the resulting parafermionic zero modes come with larger ground-state degeneracy and enable universal topological quantum operations [166,167]. Moreover, the 2D architecture of a quantum Hall system would give more freedom to implement braiding operations. There can be roughly three different approaches to realize a topological superconducting phase with a quantum-Hall–superconducting hybrid system. The first method is to induce s-wave superconductivity into the whole quantum Hall phase [168,169]. Two Majorana edge states degenerated in a single chiral edge state can be split by proximity induced superconductivity. Eventually one pair of chiral Majorana edge states gets annihilated in the bulk and only the other pair remains at the edge. The second way is to couple two counter-propagating quantum-Hall edge states via an s-wave superconducting gap and open a topological superconducting gap [170]. A third way is to couple the graphene zero Landau level of interaction-driven magnetic ordering, such as canted antiferomagnetism, with the s-wave superconductivity [171]. Despite the lure of the potential applications, the experimental progress of topological superconductivity in quantum-Hall–superconductor systems is still at an early stage. Only recently has Josephson coupling mediated by quantum-Hall edge states been demonstrated in BN encapsulated graphene with MoRe edge contacts (Figure 18c) [53]. While the differential resistance measured with a finite DC bias current above the Josephson current (black line of Figure 18d) showed plateaus, which were quantized close to the expected values of $h/(\nu e^2)$, the zero-bias differential resistance (grey line of Figure 18d) showed occasional suppressions and indicated the existence of Josephson coupling. Andreev pairs of electrons and holes propagating at different edges of the graphene are coherently connected via Andreev edge states formed along the superconducting contacts (Figure 18e). This scenario agrees with a periodic SQUID-like modulation of the Josephson current as a function of the magnetic field, which signifies a dominant Josephson current flow along the graphene edges. The coupling of counter-propagating quantum-Hall edge states through the superconducting gap was also demonstrated in graphene contacted with a narrow NbN ($H_{c2}$ > 14 T) electrode (Figure 18f) [81]. Such coupling lets the incident electron from the upstream edge state convert into a hole in the downstream edge state with a Cooper pair being injected into the superconducting drain electrode. This process is known as the crossed Andreev conversion in a sense that incoming electrons and outgoing holes reside in different edge states. Since the hole is missing in the electron of positive potential, the downstream edge state exhibits a negative potential, as shown in Figure 18g. The configuration for measuring downstream potential in Figure 18f essentially corresponds to the nonlocal setup where the bias current does not flow between the voltage probes. In this nonlocal setup, one can avoid trivial local signals stemming from the modulating density of states of graphene in the quantum Hall regime and correctly quantifies the efficiency of the Andreev process. To substantiate the scenario of crossed Andreev conversion, the negative downstream potential was shown to disappear when the temperature exceeded the critical temperature of NbN, when the energy of the incident electron exceeded the superconducting gap of NbN, or when the width of NbN electrode exceeded the superconducting coherence length of NbN. Surprisingly, spin polarized $\nu = 1$ quantum-Hall edge states were also coupled via the s-wave superconducting gap. The spin-orbit coupling inherited from the NbN superconducting electrode was suggested to enable a coupling of the spin polarized edge states. This spinless quantum-Hall edge state is of particular interest for hosting Majorana zero modes. Although this study did not demonstrate the peculiar nature of topological superconductivity, further studies on the $4\pi$-periodicity of JJs mediated by Majorana modes or the superconducting coupling of fractional quantum Hall states would be interesting and are highly demanded.

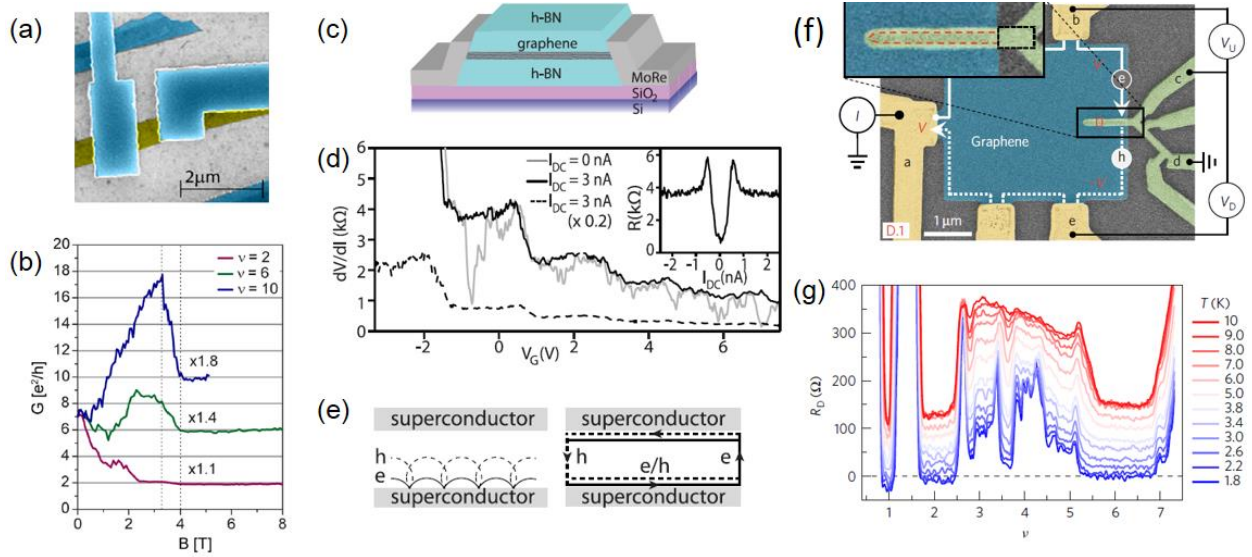

Figure 18. (a) Scanning electron microscopic (SEM) picture of Nb/graphene/Nb device. (b) Two-probe conductance at different filling factors $\nu$ [56]. (c) Schematic of MoRe/graphene/MoRe JJ. (d) Backgate voltage ($V_G$) dependence of the differential resistance (d$V$/d$I$) at different DC bias currents. Inset shows the DC bias current dependence of d$V$/d$I$ at a given $V_G$. (e) (Left panel) Quasi-classical picture of the Andreev edge mode. (Right panel) Quantum mechanical picture of the coherent Andreev edge mode and the propagation of an Andreev electron and hole pair [53]. (f) False-colored SEM picture of graphene/NbN device. Yellow, blue, and green represent the gold electrode, graphene, and NbN electrode, respectively. (g) Downstream resistance ($R_D = V_D/I$) as a function of $\nu$ at different temperatures [81].

## 5. Conclusions and outlook

Within the last decade, a vast variety of graphene–superconductor hybrid structures have been explored due to advances in nano-fabrication techniques and an enhanced understanding of mesoscopic transport phenomena. Early experimental studies relied on rather extrinsic properties of graphene such as electrostatic gate tunability or transparent superconducting contacts to expand the knowledge on mesoscopic superconducting heterostructures. However, recent developments of fabrication techniques with van der Waals materials has made the intrinsic properties of high quality graphene more accessible to studying novel physics which interconnect superconductivity and relativity. Moreover, superconducting proximitized graphene is a playground to investigate Dirac electronic optics with superconducting correlations, topological superconductivity, 2D quantum phase transitions, quantum electronic devices for quantum computation and information, and so on.

Although it was not much discussed in this review, there are active fields of application research using graphene-superconductor heterostructures. For example, graphene with its record-low heat capacity at low temperature [172] can serve as a sensitive bolometer for detecting photons absorbed in graphene. Hybrid superconducting structures have been suggested to provide fast and reliable photon detection schemes [173-175]. Recent theoretical modeling work [176] has suggested the use of properly optimized GJJs as broadband single photon detectors for the range of infrared to microwave. Graphene was also proposed to construct perfect Cooper pair splitters by exploiting its vanishing density of states near the Dirac point [177-182], although much experimental efforts are required to make the Fermi energy fluctuations much smaller than the superconducting gap. Recent experimental efforts have been made to realize graphene-based Cooper pair splitters equipped with quantum dots [183,184].

Various kinds of electronic phases of graphene have been considered in the combination of proximity superconductivity. For example, induced ferromagnetism or spin-orbit coupling in graphene was demonstrated by placing graphene onto a substrate of strong spin-orbit coupling [185-188]. This can open up research directions which combine magnetic interactions and superconductivity, such as $\pi$-JJs [189-192] or spin-sensitive Andreev reflections [181,193-196]. Once the graphene edge can be controlled to be either zigzag or armchair, the role of the valley symmetry of the time-reversed Cooper pairs can be investigated and open up the possibility of combining superconductivity and valleytronics [159,197-199]. When graphene is mechanically strained, a pseudomagnetic field can be induced due to the change of hopping amplitude to the neighboring atoms [26,200]. In this situation, electrons in different valleys feel opposite directions of the magnetic field. Strained graphene gives the unique opportunity to study proximity-induced superconductivity under strong magnetic field without suffering from degradation of the superconductivity [29-33].

## Acknowledgments

This work was supported by the National Research Foundation (NRF) of Korea through the Science Research Center (SRC) for Topological Matter, POSTECH, Korea (Grant No. 2011-0030046 for HJL). GHL acknowledges generous academic and financial supports by Philip Kim during the period of preparation of this review.